\documentclass[acmsmall]{acmart}

\def\BibTeX{{\rm B\kern-.05em{\sc i\kern-.025em b}\kern-.08emT\kern-.1667em\lower.7ex\hbox{E}\kern-.125emX}}
    
\copyrightyear{2020}
\acmYear{2020}
\setcopyright{acmlicensed}
\acmConference[Woodstock '18]{Woodstock '18: ACM Symposium on Neural Gaze Detection}{June 03--05, 2018}{Woodstock, NY}
\acmBooktitle{Woodstock '18: ACM Symposium on Neural Gaze Detection, June 03--05, 2018, Woodstock, NY}
\acmPrice{15.00}
\acmDOI{10.1145/1122445.1122456}
\acmISBN{978-1-4503-9999-9/18/06}

\newcommand{\etal}{{\em et~al.}}
\usepackage{graphicx}
\usepackage[linesnumbered,ruled,vlined]{algorithm2e}
\usepackage{comment}
\usepackage{url}
\usepackage{listings}
\usepackage{subfig}
\usepackage{todonotes}
\usepackage{csquotes}
\usepackage[official]{eurosym}

\definecolor{beaublue}{rgb}{0.74, 0.83, 0.9}

\begin{document}

%
% The "title" command has an optional parameter, allowing the author to define a "short title" to be used in page headers.
\textcolor{blue}{\title{Self-Organizing Teams in Online Work Settings}}

%
% The "author" command and its associated commands are used to define the authors and their affiliations.
% Of note is the shared affiliation of the first two authors, and the "authornote" and "authornotemark" commands
% used to denote shared contribution to the research.
% \begin{comment}
 \author{Ioanna Lykourentzou}
 \affiliation{%
   \institution{Utrecht University}
   \country{The Netherlands}}
 \email{i.lykourentzou@uu.nl}

 \author{Federica Lucia Vinella}
 \affiliation{%
   \institution{Utrecht University}
   \country{The Netherlands}}
 \email{f.l.vinella@uu.nl}

 \author{Faez Ahmed}
 \affiliation{%
   \institution{Massachusetts Institute of Technology}
   \country{USA}}
 \email{faez@mit.edu}

 \author{Costas Papastathis}
 \affiliation{%
   \institution{University of Peloponnese}
   \country{Greece}}
 \email{cst12079@uop.gr}
 
 \author{Konstantinos Papangelis}
 \affiliation{%
   \institution{Rochester Institute of Technology}
   \country{USA}}
 \email{KPapangelis@me.com}
 
 \author{Vassilis-Javed Khan}
 \affiliation{%
   \institution{Eindhoven University of Technology}
   \country{The Netherlands}}
 \email{v.j.khan@tue.nl}
 
 \author{Judith Masthoff}
 \affiliation{%
   \institution{Utrecht University}
   \country{The Netherlands}}
 \email{j.f.m.masthoff@uu.nl}
 
% \end{comment}

%
% By default, the full list of authors will be used in the page headers. Often, this list is too long, and will overlap
% other information printed in the page headers. This command allows the author to define a more concise list
% of authors' names for this purpose.
%\renewcommand{\shortauthors}{Lykourentzou et al.}

%
% The abstract is a short summary of the work to be presented in the article.
\begin{abstract}
As the volume and complexity of distributed online work increases, the collaboration among people who have never worked together in the past is becoming increasingly necessary. Recent research has proposed algorithms to maximize the performance of such 
teams 
by grouping workers according to a set of predefined decision criteria. This approach micro-manages workers, who have no say in the team formation process. Depriving users of 
control over who they will work with
stifles creativity, causes psychological discomfort and results in less-than-optimal collaboration results. In this work, we propose an alternative model, called Self-Organizing Teams (SOTs), which relies on the crowd of online workers itself to organize into effective teams. Supported but not guided by an algorithm, SOTs are a new human-centered computational structure, which enables participants to control, correct and guide the output of their collaboration as a collective.
Experimental results, comparing SOTs to two benchmarks that do not offer user agency over the collaboration, reveal that participants in the SOTs condition produce results of higher quality and report higher teamwork satisfaction.
We also find that, similarly to machine learning-based self-organization, human SOTs exhibit emergent collective properties, including the presence of an objective function and the tendency to form more distinct clusters of compatible teammates. 
%This work among the first to explore self-organization in the context of distributed team formation, and we use its results to draw useful conclusions about the future of online collaborative work.
\end{abstract}



%
% Keywords. The author(s) should pick words that accurately describe the work being
% presented. Separate the keywords with commas.

%
% A "teaser" image appears between the author and affiliation information and the body 
% of the document, and typically spans the page. 
%%  \includegraphics[width=\textwidth]{sampleteaser}
%%  \caption{Seattle Mariners at Spring Training, 2010.}
%%  \Description{Enjoying the baseball game from the third-base seats. Ichiro Suzuki preparing to bat.}

%
% This command processes the author and affiliation and title information and builds
% the first part of the formatted document.
\maketitle

\section{Introduction}
As online work increases in complexity, crowdsourcing research and practice turns more and more into collaboration. Examples of problems where large-scale crowd collaboration has proven valuable include scientific research and article authoring~\cite{vaish2017crowd}, designing software prototypes~\cite{Retelny:2014:ECF:2642918.2647409}, writing stories~\cite{Kim:2014:EEC:2531602.2531638, kim2017mechanical}, and collaborative idea generation~\cite{Siangliulue:2016:IIL:2984511.2984578}. Crowd teams differ from traditional teams working in face-to-face or in online corporate settings in that they (i) consist largely of people who have not worked together before prior to the crowdsourcing task, (ii) need to perform efficiently in relatively little time, and (iii) cannot be assumed to share a common set of workplace values, such as loyalty to a specific organization. 

The scale of crowdsourcing and its online nature, means that algorithms are often involved when it comes to forming the crowd teams. This approach is in contrast to typical team formation taking place in face-to-face settings, where a manager or an expert user may assign team members to the team based on a knowledge of their skills, or of their past history of working together ~\cite{gorla2004should}.

Algorithm-based methods for team formation attempt to do the same at scale, either before the task starts, by pre-profiling workers and then assigning them to teams~\cite{Rahman:2019:OGF:3311351.3311362}, or during the task, by changing the group synthesis to improve collaboration elements such as viewpoint diversity~\cite{Salehi:2018:HCD:3290265.3274420} or interpersonal compatibility ~\cite{lykourentzou2016team}. Broadly speaking, team formation algorithms belong to the category of crowdsourcing management algorithms, the objective of which is to match persons to other persons or persons to tasks, and in this way to optimize the speed or efficiency of crowdsourced task production. 

The problem with existing algorithms is that they \textbf{micro-manage workers by design}. Workers have little to no say in who they collaborate with, for how long or how. In the case of algorithms that assign individuals to teams \textit{before the task} starts, the algorithm deduces the individual's performance within their future team based on a set of pre-calculated profiling features, such as expected performance. Unfortunately, these features are often incomplete and subject to significant volumes of noise, given the sparsity of data in regards to - for example - complex skills. In addition, elements such as interpersonal compatibility, which are inevitably only revealed after the team has worked together, are not taken into account. If the collaboration does not go as expected, workers cannot signal so, and they cannot change teams during the process. In the case of algorithms that do change the teams \textit{during the task}, these changes happen again by deduction and in an invasive manner, without explicitly asking the workers or requesting feedback from them. Workers literally receive a message informing them that they have been placed with a new team, and they have to adapt to this decision. User control of the collaboration process is largely absent, with repercussions that can range from psychological discomfort ~\cite{rasmussen2006teamwork} 
to less-than-optimal collaboration results ~\cite{de2001minority}.  

As one may expect, not actively involving the workers, but rather assigning them directly to a group (or a task) comes with disadvantages. Latest research in management sciences~\cite{LawlerIII2006} and also crowdsourcing~\cite{Retelny:2017:NWE:3171581.3134724} indicates that too close a monitoring can stifle worker creativity and initiative-taking: two features that are absolutely necessary in creative, complex teamwork. In parallel, a substantial body of literature on the nature of collaboration indicates that providing agency to the workers can result in better collaboration and increased feelings of control ~\cite{haas2016secrets}. Finally, trusting the workers to co-design the task workflow, which in the case of collaborative work means to co-decide who works with whom, has proven to be beneficial for solving complex problems, for which no evident solution is to be found ~\cite{costa2018trust}.  

A problem therefore arises: (how) can we balance the necessity for algorithm-based team formation, which is necessary given the scale of crowdsourcing, with the need to give each individual online worker a say in who they will work with to best accomplish the task? 
In this work we explore a new concept: \textbf{algorithm-assisted self-organization}; an approach designed to empower 
%crowd% 
online workers with the opportunity to choose their teammates, and thus ``guide'' the algorithmic process of team formation. Self-organization is a management principle that has often been used in other types of collaboration settings, such as Open Source Software communities and agile corporate teams~\cite{Beck2001}. It has however never been used in an online collaboration setting that requires the parallel involvement of a team formation algorithm. \textit{Self-Organizing Teams (SOTs)}, which are the result of our approach, 
\textit{can be better understood through a metaphor from the machine learning field}. In a similar way that Self-Organized Maps (SOMs) gradually re-arrange their input data
to form highly coherent clusters based on Euclidean distances, people in the SOTs concept gradually discover the best teammate(s) to work with, based on reputation and personal experience of working together as the task progresses. An algorithm assists this process, by using people's explicit teammate preferences to form teams that are dictated not by an assumption of which teams could work, but based on which teams the workers indicate that will work. The fact that the algorithm does not rely on any pre-established assumption, e.g. which worker features to use to form the teams, means that SOTs are largely task-agnostic and can thus be applied on a variety of complex 
online work tasks. Being task-agnostic is the second advantage of this method, given that currently complex task workflows tend to be both expensive to develop and over-adapted to a specific task~\cite{doi:10.1111/j.1467-9310.2010.00596.x}.

Self-organization in crowdsourcing is a new research line, and many alternatives can be explored regarding its implementation. In this paper we work with a collaborative-competitive setting, where the task is accomplished in discrete steps, called rounds. The specific task that we work with is collaborative short story writing, chosen as it is a complex task that does not require specialised expertise, does not have one evident solution, and can highly benefit from co-creation ~\cite{dobao2013collaborative}.
During each round, workers collaborate in small teams to continue a main story. 

At the end of each round workers decide as a collective, on a single ``winning'' team, by voting their preferred story continuation. Before the next round starts, they can opt to stay with the same team or to change, and an algorithm accommodates this process based on their choices. Gradually, round by round, workers get to explore the ``space'' of candidate teammates to discover those with whom they might work best to win. We juxtapose this setting with a setting where the crowd teams remain fixed throughout the process, without any choice for self-organization or agency. Results shed light into how people select their teammates and how they gradually form desirable team ``clusters'', the effect that self-organization has on their collaboration, but also on their self-perceived effectiveness, sense of control and in general on the way that they decide. 

The rest of this paper is organized as follows. In the next section we present related literature, including the limitations of current algorithm-driven team formation. We also summarize related work insights about self-organization from other domains, and present the hypotheses of this study. Next, we go through the study design, including the self-organizing team framework and its supporting SOT algorithm. Then we describe the experimental process and the main experimental results. Finally, we discuss the study findings, with special emphasis on future directions given the novelty of this approach, and conclude the paper.

\section{Related Work}
\subsection{Team Formation Algorithms: A trend to micro-manage 
workers}\label{sec:related-algorithms-micromanagement}

One may distinguish two types of algorithms guiding the online 
team formation process: i) algorithms which select which worker should work with whom \textit{before} the task begins, which we hereby call crowd~\textit{team building algorithms}, and ii) algorithms that manage the team processes \textit{after} the task has begun, which we call crowd~\textit{team coordination algorithms}. 

Team building algorithms view team formation as a mathematical optimisation problem. They have been developed for the scale of online work and crowdsourcing, which make it impossible for traditional approaches (e.g. a human manager) to put together an effective team. Assuming a large pool of workers with known profiles (e.g. skill level) and a large pool of tasks, the objective of a crowd team building algorithm is to match each task with a group of workers to accomplish the task optimally within given constraints, such as deadline, upper budget, or minimum quality. In this line, Rahman~\etal~\cite{Rahman:2019:OGF:3311351.3311362} proposes an algorithm that utilises affinity and upper critical mass to recommend tasks to groups of online workers, taking also into account the aggregated worker skills and cost of effort.  
In addition, Liu~\etal~\cite{7248382} propose a task pricing algorithm that attempts to assemble a team of crowd workers to complete a given task with the lowest cost. Both these works rely on predictive learning algorithms, which make their team formation choices based on a limited set of pre-calculated worker profiling features, without worker feedback. The risk of relying on such algorithms is to reduce workers to a set of dimensions that do not account for a person or -- since teams consist of people -- a team's evolution, intentionality, and needs~\cite{FARAJ201862,doi:10.1177/0162243915606523}, and therefore risk creating rigid, incomplete and less-than-optimal team structures.

Crowd team coordination algorithms manage the team structures during the task. In this research direction, Kim~\etal~\cite{Kim:2014:EEC:2531602.2531638} propose Ensemble, a methodology to create stories through the crowd. Ensemble is coordinated in a top-down manner, with teams that feature story ``leaders'' directing a higher-level story vision, and workers materializing this vision into concrete story pieces. Workers do not get to decide on the final story, and their contribution is limited to proposing drafts, comments and votes, i.e. assisting the leader.
Valentine~\etal~\cite{Valentine:2017:FOC:3025453.3025811} and Retelny~\etal~\cite{Retelny:2014:ECF:2642918.2647409} propose Foundry, a crowd management system that also relies on a hierarchically structured approach, to ``assemble'' workers into role-based teams.
Workers can request a change in the initial team structures, but the final decision is taken in a top-down manner by a small number of expert workers and the task requester. 
Eventually, workers are notified as to which team they pertain. Although this system does incorporate worker feedback, it does so in the form of worker suggestions and not decisions.
Salehi~\etal~\cite{Salehi:2018:HCD:3290265.3274420} propose an algorithm that rotates workers across teams based on viewpoint diversity, to improve the quality of a creative design task. Workers are not asked whether they would like to switch teams. Although the use of the rotation algorithm did produce higher quality task results, the authors acknowledge that forcing participants to work with specific people led to discomfort, and that the teams actually wished for the ability to prevent algorithm rotation decisions.
 Zhou~\etal~\cite{Zhou:2018:SDT:3173574.3173682} propose an algorithm based on multi-armed bandits with temporal constraints, which explores different team structures and timings to apply these structures on a team. The algorithm explores various exploration-exploitation trade-offs and chooses, from a finite set of possible structural changes, when and which change it should make in the structure of a team. However, in this case too, the team formation algorithm is the driver and principal decision-maker behind the team structure. 
On the analogous subject of crowd-led authored content, \citet{kim2017mechanical}'s Mechanical Novel system provides AMT crowd workers the opportunity to create short fiction stories in loops of reflection and revision, and 
in a manner that decentralizes the decision-making process much more than past systems. Although these past works, and especially \cite{Salehi:2018:HCD:3290265.3274420} and \cite{kim2017mechanical} touch on the subject of user agency, the present study explores team behavior in a setting that incorporates user control systematically, as the utmost prominent aspect of the system design.

Finally, \citet{Salehi:2017:HCS:2998181.2998300} and~\citet{lykourentzou2017team} both propose systems of automated team formation, which take worker feedback into account regarding the quality of past collaborations. In both these systems, the workers evaluate their peers \textit{after} having worked with them, and the system uses these ratings to calculate an overall benefit function that drives team formation on a new, self-contained task. Our study shares similarities with these works in that it also actively requests worker feedback regarding past teammates. 
It differs in that the objective function to be optimized is not decided a priori by the algorithm designer, but it is generated on-the-fly by the self-organisation decisions of the user collective.

Latest research~\cite{Retelny:2017:NWE:3171581.3134724}  acknowledges that externally, for example by an algorithm, pre-defined planning of the way the team will work or its structure, is not optimal, especially for open-ended, complex tasks. The reason is that such planning can inhibit workers from adapting to the needs of the complex problem they need to solve, in real-time. The authors suggest that complex problems require approaches that enable open-ended adaptation. This work is in line with prior research on \textit{accountable governance} work models, which showcases that creativity is fostered when individuals and teams have relatively high autonomy in their everyday work processes, as well as a sense of ownership and control over their own work and ideas~\cite{andriopoulos2001determinants,amabile2018creativity}. Allowing independence around work processes also enables workers to resolve and adapt to problems in ways that better utilize their expertise and creative thinking skills~\cite{amabile1998kill}.

SOTs is an approach in this direction. In contrast to the centralised manner of organizing crowd work, SOTs enable workers to collectively decide, rather than suggest, on the best course of action as the task progresses, flexibly adapting both the involved team structures and the output solution as the task unfolds. 

\subsection{Towards giving workers team formation agency}

User agency has been studied under at least two major approaches in the literature: as a product of direct user negotiation (or reciprocal agreement), and as a product of algorithmic mediation. The first approach demands mutual agreement among the workers before forming teams. The second approach uses an algorithm to mediate this process and determine the teams based on preferences.

The problem of agency as the product of user negotiation has been investigated through agent-based modeling by research like the one of \citet{guimera2005team}. Their work simulates the emergence of collaboration networks in creative enterprises based on the users' propensity to collaborate under multiple constraints (team size, the fraction of newcomers in new productions, and incumbents' tendency to repeat previous collaborations). 
The explicit intent to remain in a team -- ergo, their direct negotiation -- is part of a study on self-assembled teams by~\citet{zhu2013motivations}. The study enabled online gamers to join or to leave virtual teams across some period. The players could only join teams sequentially, and their decision to remain in those was determined by: a) whether they played together synchronously, b) whether the team did not change in size during the cooperation, and c) whether the team became inactive for longer than 30 minutes (after which the team was dismantled).
The study of~\citet{tacadao2015generic} observed the emergence of team self-assembly, under the different approach of algorithmic mediation. The model in~\cite{tacadao2015generic} is designed for collaborative learning scenarios where the cohorts produced are evaluated on the parameter constraints and the number of teammate preferences they satisfy.
Finally, the work of~\citet{meulbroek2019forming} studies algorithm-supported matchmaking in the context of student teams, where the authors developed a system based on the CATME algorithm to determine the preference ranking of the students.

In our study, the algorithm supports the users' choice whilst easing the complexity of the negotiation (e.g., cognitive overload, group size, etc.), which could slow down the execution of the task and deplete the users' working memory resources. However, in contrast to the studies mentioned above that examine user agency through simulations, or in settings other than online work, we propose a solution for self-organization, which is deployed with real users and on the particular setting of online work.

On the particular area of online system design,
research has only recently started exploring the perceptions of users when it comes to choosing their teammate, if they are given the choice.
G\'{o}mez-Zar\'{a}~\etal~\cite{Gomez-Zara:2019:YLW:3290605.3300889} examine how people search and choose teammates in online platforms. Their research indicates that users search for teammates based on competence, common values, similarity in social skills and creativity levels, as well as prior familiarity. They also find that users eventually choose teammates who are well-connected, with many prior collaborators. Their study concludes that future systems should be hybrid, augmenting user agency with algorithms.   
Jahanbakhsh~\etal~\cite{Jahanbakhsh:2017:YWM:3025453.3026011} examine the perceptions of users regarding automated team formation in educational settings. Their findings reveal that although users valued the rational basis of using an algorithm to form teams, they did identify mismatches between their preferred team formation criteria and those of the algorithm, and expressed the need for having a say in the process. This study too closes with the recommendation to give users more agency in the selection of their teammates, and even advocates for constrained team self-formation, in line with earlier works in the educational domain~\cite{doi:10.1177/105256299902300503}.

\subsection{Self organization for team formation}
\subsubsection{In software teams}
Self-organization is a term often used to describe the governance of software development teams, either within a company or in Open Source Software Development communities~\cite{doi:10.1177/030981680909700108}. In fact self-organization is one of the 12 principles behind the Agile Manifesto~\cite{Beck2001}. 

In this context, self-organization is defined as a process followed by teams that manage their own workload, shift tasks based on needs and best fit, and participate in the group decision making~\cite{Highsmith2004}. The ability to self-manage has been found to significantly affect the performance of agile teams, because it ``brings the decision making authority to the operational level, thus increasing speed and accuracy of problem solving''~\cite{moe2008scrum}. It has been repeatedly found to help software development teams cope with the increased dynamism, resource control autonomy, and decentralisation, which are inherent in today's globalised environments~\cite{10.1007/978-3-540-24701-2_1}. 

Self-organizing teams have certain characteristics~\cite{Takeuchi1986}. First, they are driven by ``zero information'', where prior knowledge does not apply. This enables the team members to challenge existing knowledge status quo and have the potential to create something truly novel. Second, they exhibit autonomy, as they do not have a top-down appointed leader, but rather leadership is a property that emerges as the team members divide their roles~\cite{Hoda2010}. Third, they exhibit self-transcendence, i.e. the team pursues ambitious goals, and fourth, cross-fertilization, i.e. team members have a variety of backgrounds, viewpoints and knowledge. 

In summary, self-organization seems to improve effectiveness of development teams, by providing them the ability to innovate from zero and by enabling them to maintain the autonomy and diversity of their members backgrounds.
We use these two characteristics of self-organization in the notion of online work SOTs that we introduce. SOTs, however, have one important difference with the self-organized teams in a software development context: whereas the team members of agile teams are hand-picked and then given self-organized autonomy, we apply self-organization from the team formation stage, allowing individuals to explore the ``space'' of candidate teammates available to them and discover those with whom they might work best, through a trial-and-error approach.  

\subsubsection{In complex systems theory and machine learning - the property of emergence}

Self-organization in complex systems can be defined as the emergence of global structures out of local interactions in a process that is collective, parallel and distributed ~\cite{heylighen2008complexity}. 

Emergence is a property of complex dynamical systems where global processes cannot be reduced to the sum of the fundamental units and which are generated without the need to tune control parameters to precise values. 
Self-organization, as a mechanism, is seen to emerge in numerous contexts, from the ants' food foraging, molecules formation, collective dynamics of flocks of birds and schools of fish.

Similarly, self-organization in complex dynamic systems is a principle applied in computer science and engineering by relying on autonomous entities to achieve robustness and adaptability of the simulated environments ~\cite{serugendo2003self}. Agent-based models (ABM) are micro-scale models~\cite{gustafsson2010consistent} that simulate the appearance of complex phenomena with the concurrence of micro-scale behaviours and macro-scale states. 
Nature inspired algorithms with ABM can be seen, for instance, in the particle swarm optimisation simulating the flocking of birds ~\cite{kennedy2010particle}. 
Another known example of self-organization- inspired computation is the artificial neural network of the Self organizing Maps (SOM) trained by unsupervised learning. SOMS identify spatial-temporal patterns in complex systems by the use of dimensionality reduction and clustering of observations based on similarity. On the topological level of social networks, some units exhibit a tendency for optimising their utility by clustering together with other units. The clustering, when based on preferential choice, is observable in a number of emerging social patterns like the variation of the classic Schelling Segregation model with Strategic Agents~\cite{chauhan2018schelling}.
The property of emergence will be extensively explored in this paper within the practice of Self organized teams formation and coordination, in particular under the condition of complex macro tasking and pair-wise matching.

\subsubsection{In gaming and online dating settings} Matching users in multiplayer online games has been investigated from the perspective of multiple fields, from networking \cite{agarwal2009matchmaking} to algorithmic optimization \cite{farnham2009method}. Most of these studies focus on predictive feature-based matchmaking, which accounts for the skill levels of the players~\cite{minka2018trueskill}, their personalities~\cite{chan2018keeping}, their play style~\cite{farnham2009method}, their ranking~\cite{graepel2006ranking}, or an ensemble of the above~\cite{riegelsberger2007personality}. There is, however, a gap in the related work around the concept of self-organization for online virtual gaming. The most pertinent investigation of self-organization comes from analyzing the network structures of popular social network gaming platforms such as Steam~\cite{becker2012analysis} or 100.io~\cite{schiller2018inside}. Platforms of this kind take advantage of both the pre-existing social networks of the players, as well as the use of lobbies. Steam's peer-to-peer matchmaking is built around the concept of a lobby that can be queried by the ISteamMatchmaking API interface~\cite{matchmaking}. Skill-based matchmaking is added on top of this system, while players can search other available players based on their geographic proximity. Social networking is a strong component of gaming platforms as they provide the option to play with friends and friends of friends from the accounts list. On the contrary, crowdsourcing platforms do not supply the crowd or the requester with support to build social ties. This lack of networking reinforces the need for user agency in crowd matchmaking to stimulate the workers' engagement and satisfaction~\cite{lykourentzou2017team}.

In online dating settings, collaborative filtering recommender systems have been widely used to limit information overload \cite{brozovsky2007recommender}, propose best matches, and provide some level of serendipity. Although engineered to fit with the concept of user-centered systems, popular dating apps have been accused of reinforcing popularity biases~\cite{rudder2014dataclysm}, and to intrinsically disfavor minority groups~\cite{birger2015date}. In our study, we deliberately remove some of the most sensitive user details, such as the workers' real names and their profile pictures, to lower the number of user choices that are based on cognitive and social biases. We do not use recommender system methods in the current framework, since the pool of workers is limited to a manageable size. Our study on self-organization of smaller batches of workers collaborating in dyads follows the upward trending research on the ``Renaissance of small groups'' as seen in studies dated from 2018 onward~\cite{harris2019joining}.

\subsubsection{In crowd team settings} 

Lykourentzou \etal \cite{lykourentzou2016team} explores how to dynamically create crowd
teams from a large population of potential workers without any prior knowledge of worker profiles, which they call team dating. The idea behind team dating is that the task authors delegate team building to the crowd workers themselves and ask them to try out different candidate co-workers, evaluate them, indicate those that they like working with, and make crowd teams based on these indications.
Rokickiet al.~\cite{Rokicki:2015:GTC:2736277.2741097}  also talk about self-organization for crowd team building, by studying various team competition strategies including balanced teams, self-organizing teams (built upon one worker as the first administrator, accepting/denying the contribution of other members) and a combination of team and individual strategies. The results show that teams outperform individuals at task annotation without impacting the quality of the end product. In this paper we explore in particular the effects that self-organized dyadic team formation has on the individual's sense of entitlement, reward and creative outreach, without the presence of a single-handed administrator to moderate and steer the team towards effectiveness. 

Some works have started exploring the notion of exploration-exploitation in crowd work settings, albeit with a different meaning. 
Johari~\etal~\cite{DBLP:journals/corr/abs-1809-06937} consider the exploration-exploitation trade-off in the context of labor platforms where flash teams and on-demand tasks can be improved by the assistance of a matching algorithm modelled on a binary classification of the agents in a pair-wise fashion. The exploration herein is defined as the learning performance of untested teams against the exploitation of repeating previously tested teams. The stylised model relies on known and unknown features for near-optimal matching in consideration to the population distribution (apriori knowledge) and the payoff structure under the aggregated performance objective with the lowest regret. Unlike Johari et al., we present a self-organized team framework which does not reduce the workers to simplistic terms (analogous to binomial label classification problems) and does not preclude existing knowledge of the performance of the collaborators prior the start of the job. More so, the intensification and diversification of the strategy is not led centrally by a coordinating algorithm, but defined by the worker's initiative only, as we allow the collaborators to guide the systemic changes occurring across the work phases.

Another application of the exploration and exploitation trade-offs in the search space is presented by Xiang et al. ~\cite{feng2019social} with a mathematical model of social team building optimization (STBO) and the adoption of swarm theory and swarm intelligence. According to this research, the team-building states, or phases, lead to a converging point dependent on the exploration of new solutions and the exploitation of already visited neighbourhoods considering the energy and entropy of the team. This team building optimisation algorithm greatly depends on a defined social team hierarchy which is not a prerequisite in the model we propose, enabling the process of team formation to functionally generate a hierarchy as a result.  

In Zhou~\etal~\cite{Zhou:2018:SDT:3173574.3173682} exploration-exploitation means changing the team structure or keeping the same, and it is decided by an algorithm, based on a finite set of (five) decision elements. In our case it means changing teammate or staying with the same, and it is decided by users, based on an unknown set of decision elements, constrained only by the number of team dynamics' cues they can process through their text-based interaction with another user. Humans in this case take up the role of exploring the decision space, instead of an algorithm.

\subsection{Lessons learned from the literature and contributions of this work}
As we saw in the previous sections, current approaches in algorithm-based large-scale team formation are primarily top-down, and they do not allow workers freedom in regards to who they will collaborate with, or decide on the team work structures and the collective task outcome. Related literature also highlighted the fact that such top-down, micro-management techniques may work for routine, very well defined tasks, but they fail for ill-defined, complex and creative work, such as innovation generation or creative writing. In contrast, bottom-up approaches, have been found to promote creativity and team performance in complex task settings, because they motivate workers to own and take responsibility for the creative process and its outcome.

Aiming to address the above, in this work we will explore, for the first time a \textit{new human-centered computational structure}, namely Self Organizing Teams (SOTs). The SOT structure does not just give users agency over who they will work with, or help them form these teams at-scale using an algorithm. It also enables users to control, correct and guide the task output solution as a collective, by competitively filtering out weak candidate solutions and selecting the most promising one, as the task progresses. Through the self-organization and collective control over the task solution -- and supported but not guided by an algorithm -- users gradually build a consensus-based and community-approved global solution.

\section{Methodology}

\subsection{Reward}\label{reward}  
To motivate users to win and consider switching teams
if needed, the workflow includes an element of competition, in the form of a bonus payment (for paid workers) or an increase of the obtained score (for non-paid users). More specifically, every time a team
wins, its members gain an extra amount equal to the base pay (for paid participants) or the base score (for non-paid participants). 
In line with the latest recommendations for academic requesters regarding fair payment\footnote{\url{https://wearedynamo.fandom.com/wiki/Guidelines_for_Academic_Requesters}}, the base pay for Amazon Mechanical Turk (AMT) workers was \EUR{5} for a total task time of approximately 30 minutes. Given the three rounds (three chances to win), worker remuneration could reach a maximum of \EUR{20}. For non-paid participants, the monetary payment was transformed to a base score; these participants would get a base score of 5 points for participating in the task and 5 more points every time their team won. As a further element of driving competition among non-paid participants, a leaderboard was shown at the end of the task, illustrating their placement relative to the other participants.
As we will discuss later in the results and in the Discussion section, this reward choice directly affects the teammate selection decisions made by the participants throughout the task, i.e. it determines to a large extent the objective function of the SOT as a collective.

\subsection{Group formation requirements}
Group size is critical to team performance, especially when it comes to creative tasks, with research showing that the number of creative ideas per person increases as team size decreases~\cite{bouchard1970size,renzulli1974fluency}. 
Thornburg further shows that a group's Creative Production Percent (the percent performance of a group compared to the performance of an individual) improves as the team size decreases, until it reaches its peak at a group size of two, i.e. dyads. The reason,
is that dyads have a unique one-to-one capability to share and exchange ideas, while the inhibitors that typically occur in larger teams, like social loafing, groupthink and production bottlenecks, are less likely to occur in these groups~\cite{brajkovich2003executive, knights1994managers}. Furthermore, dyad interactions permit the observation of key team processes like coalition formation, inclusion/exclusion, power balances and imbalances, leadership and followership, cohesiveness, and performance~\cite{williams2010dyads}, all of which have been linked with expressions of team member agency and autonomy in various settings~\cite{fausing2013moderators,cordery2010impact,hoegl2006autonomy, langfred2000paradox}. 
Taking the above into account, in this study we work with dyads. 
We elaborate how our model can be extended to accommodate triads or larger groups in the Discussion section \ref{from_dyads_to_teams}.

In addition to team size, it is also important to decide the size of the batch, i.e. how many people will be recruited for a single SOT lifecycle. The batch size needed to be an even number, so that all participants find a team to join. Given that we work with dyads across three rounds, the minimum batch size for a meaningful choice is 6 (less than that means that participants are highly likely to end up with a teammate they do not prefer). Then, the larger the batch size, the more options a participant would have in selecting candidate teammates. However, too large a batch means that people will not be able to process and compare effectively all candidate teammates, due to both short-term memory limitations but also due to the limited time they have to choose a teammate. With the above in mind, we opted for batches of 6 to 12 people. This allows for an adequate number of different dyadic team formations, while keeping the cognitive load of processing multiple user profiles manageable ~\cite{knijnenburg2012explaining, bollen2010understanding}. 

\subsection{Task description}\label{task_description}
\subsubsection{Task requirements}
In defining the appropriate task for the study, we needed to take into account a number of requirements. First, we needed a task that involves complex, open-ended work for which no single solution is evident, cannot be easily decomposed to fixed workflow structures, and requires workers to maintain the global context and full semantic overview of the
problem while iteratively refining it~\cite{altshuller1999innovation, majchrzak2013towards}. Recent crowdsourcing literature refers to these tasks as macrotasks~\cite{schmitz2018online, lykourentzou2019macrotask}, distinguishing them from microtask-based work. Examples of candidate macrotasks included brainstorming, writing, prototyping, product development, innovation development, formulating an R\&D approach and so on. These macrotasks require the combination, through trial-and-error, of the diverse knowledge, skills and creativity of multiple collaborating individuals ~\cite{lykourentzou2019macrotask}. As such, these tasks can benefit the most from the SOTs structure, the purpose of which is precisely to enable the continuous ad-hoc adaptation of the solution output and work processes to the task needs. On the other hand, tasks that are close-ended, those with known knowledge and skill interdependencies~\cite{argote1982input}, or tasks for which a specific work process can be determined a priori~\cite{okhuysen200910}, would not be appropriate candidates, as these can be optimally solved through workflow management and crowd coordination algorithms, like the ones described in Section~\ref{sec:related-algorithms-micromanagement}.
Furthermore, the task needed to adhere to three key criteria for an online creative work setting, involving people working online together for the first time: no requirement of prior expertise, short duration, and ability to express creativity~\cite{dow2011prototyping}. 

The task that was selected to fulfill the above criteria is a creative writing challenge, inspired by the exquisite corpse method \cite{brotchie1995surrealistgames}, where participant teams co-create a fictional story by gradually building on each others' contributions, across multiple rounds. Creative writing tasks of the above type, can be used for applications such as rapid game scenario design (e.g. to provide more truthfulness and content to online gaming AI) or to generate content for the creative industries (film making, advertisement, etc.). In line with the SOT framework, the task allows for cycles of collaboration, where teams work internally to produce candidate story continuations, and competition, teams compete for the single best story continuation through peer review. 

The pre-authored story used as input to the creative writing task of this study is the following:

\begin{quote}
{\fontfamily{qcr}\selectfont
At a restaurant, Mary receives an SMS and reads the following message: ``Your life is in danger. Say nothing to anyone. You must leave the city immediately and never return. Repeat: say nothing.'' Mary thinks for a second and then \ldots
}
\end{quote}

\subsubsection{Timing} The proposed framework is designed to work with an ongoing flow of users joining the task at slightly different times, as it is typically the case when working with commercial crowdsourcing platforms. The system is programmed to account for a minimum threshold of registrations (between 8 and 12, depending on the flow of the workers) and a maximum waiting time, after which it redirects the workers to unique batches of experiments. By monitoring and assessing the registration flow of the workers across multiple trial runs, we were able to determine the average batch size for the experiments without encountering critical levels of delays that could overtake a large portion of the task. Even though the job fitted some of the characteristics of micro-tasks  (real-time, short-termed, and unique), to be able to hire workers from the Amazon Mechanical Turk platform, its core is designed to be executed as a macro-task (complex, collaborative, and analytical).

\subsection{Study Setting}
Commercial crowdsourcing platforms do not encourage collaboration and do not permit worker allocation in self-organized teams. For this reason, we designed a tailor-made framework and its supporting platform. The proposed work has been designed with the intention to address the individual's ability to leverage collaboration with a certain degree of freedom and given creative agency as structural part of the collaboration process. 
\subsection{Overview of the Self-Organizing Team Framework}
\begin{figure}
    \centering
   \includegraphics[width=1\textwidth]{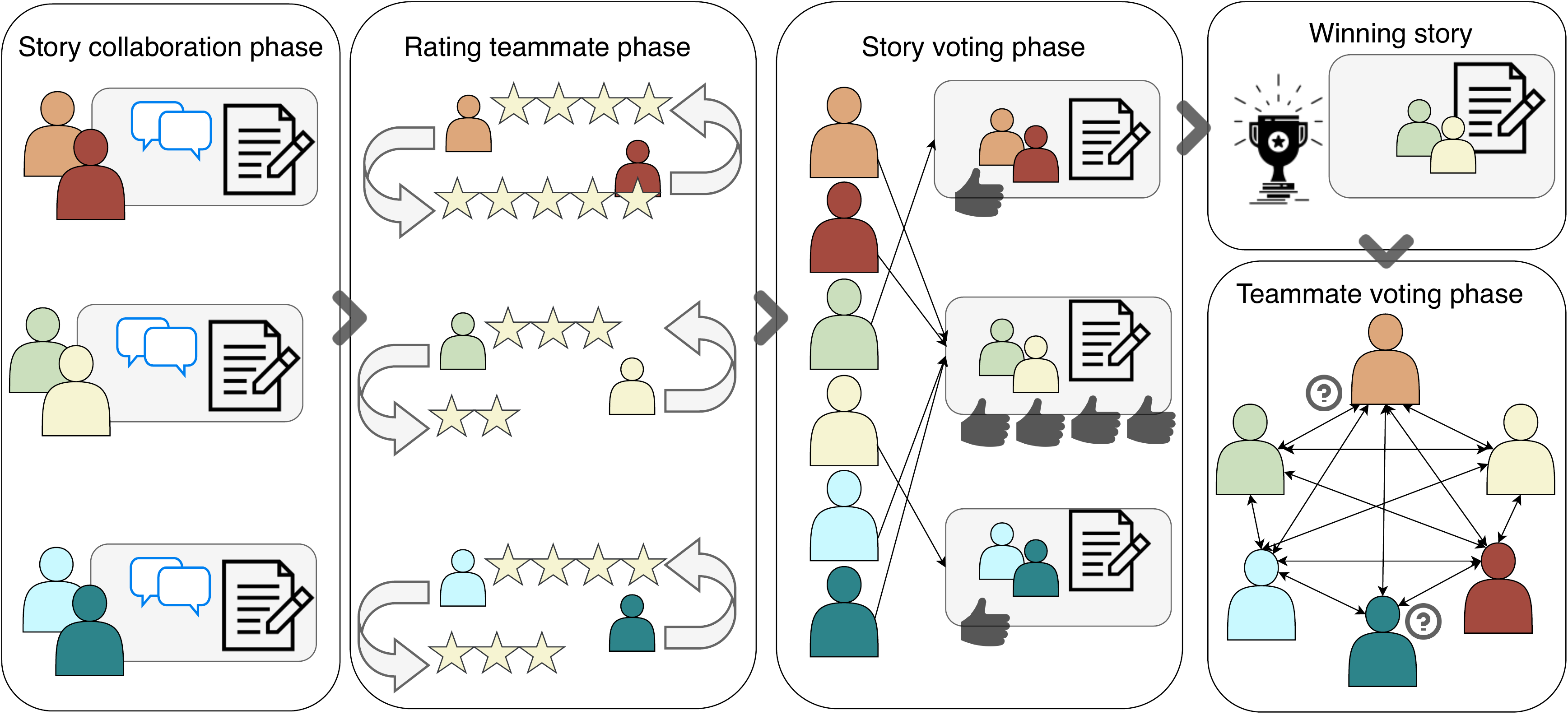}
    \caption{The SOT collaborative-competitive framework. Participants \textit{collaborate} in dyads to progress a creative task, which for this study is continuing a short fictional story, and then they evaluate their in-between collaboration (story collaboration and rating teammate phases). Next, teams \textit{compete} for the best story continuation through peer review (story voting and winning story phases). Finally, participants indicate which teammates they want to collaborate within the next round (team voting phase). The SOT algorithm forms the teams based on these choices, facilitating self-organization. The process repeats for three rounds.}
    \label{fig:3_phases}
\end{figure}

The SOT framework presented below is illustrated in Fig.~\ref{fig:3_phases}. It is designed to function on the basis of a collaborative-competitive setting, completed across various rounds. During each round (collaboration phase) workers form teams 
and collaborate with their teammates to progress a creative task, which in our case is the continuation of a -- pre-authored and same for all teams -- short fictional story. The task is described in more detail in Section~\ref{task_description}. At the end of each round (competition phase), users employ peer review to vote for their favorite story continuation, without the possibility to vote for their own team's story. The winning story is appended to the main story, the winning team is announced and its members receive an award, which for paid crowd workers is monetary and for non-paid volunteer participants is in the form of a score. Then, before a new round starts, users decide whether they wish to continue with their previous team, or change. The \textit{SOT algorithm} receives this input and forms the teams, aiming to accommodate to the best possible extent the teammate preferences of each user in the batch. The algorithm is described in detail in Section~\ref{sot_algorithm}. The cycle of collaboration and competition continues, with users each time assessing the benefits and risks of switching teammates (e.g. higher probability of winning in case of switching teammates in an under-performing current team, but a steeper curve for learning to work together), making their team formation decisions, forming new or old teams and continuing the main story as it was formed in the previous round. At the end of the final round, users are presented with the outcome of their collective work (finalised main story), and a ranked user list showing the number of times each user won, in descending order. Finally, users fill in a final questionnaire on their experience.

\subsubsection{Task Initialization: Recruiting Participants} 
Users register to our experimental platform in two ways. In the case of paid crowd workers, they enter with the credentials of the crowdsourcing platform used to hire them, in order to facilitate their automatic payment once the task finishes. In the case of volunteer participants, they register with a unique identification number. 
For each experiment of our study, once the desired \textit{batch} of people has arrived, the experimental platform stops hiring people, and those registered are moved to the next step. From that step onward, the system is synchronized, meaning that all workers are moved from one step to the other after a specific amount of time has elapsed. Users are always shown the remaining time, the round that they are currently in and the amount they have won so far on the top of their screen.  

\subsubsection{Setting up: Instructions, Demographics and Individual Creative Work Sample} 
Workers are presented with the task instructions, which briefly present the creative task, its goal and their reward upon completion. This stage takes just over a minute, and users are given the following instructions:
\begin{enumerate}
    \item \textbf{The task}: Users are instructed to work in teams 
    to continue a short story in English. They are informed that the task has 3 rounds and that in each round the story that gets the most votes wins and is appended to the main story. This extended main story will be then shown as a prompt to all participants and a new round will begin.
    \item \textbf{The goal}: Users are instructed that their goal is to be in the winning team. They are explained that in each round the system will automatically match them with the same or with another teammate. They are prompted to do their best so that your team's story gets the most votes.
    \item \textbf{The reward}: Users are explained that they get a [Base reward\footnote{The Base reward can be either a Base pay or a Base score, depending on the type of participant batch (paid or volunteer); see Section~\ref{reward} for details.}] for participating in the task and an extra [Base reward] every time they are in a winning team (3 chances for this). They are also told that their maximum gain is [4 x Base reward].
\end{enumerate}

Next, users are asked to fill in a short questionnaire about their demographic information, namely: (i) gender, (ii) age, (iii) ethnicity, (iv) education level, (v) employment status, (vi) prior experience in a creative task like the one they are about to work on, and (vii) self-perceived creativity levels.
To assess the creative self-efficacy\footnote{Creative self-efficacy is defined as individuals' beliefs in their ability to produce creative ideas. Past work~\cite{shin2012cognitive} has shown its positive relationship with creative outcome.}, we used the eight-item scale from~\cite{carmeli2007influence, chen2001validation}.
In our experiments, this stage takes less than a minute for completion.

\subsubsection{Creative Writing: Sample story writing}
Then, each user is presented with the start of a pre-authored fictional story (same for all users), and is asked to write a brief continuation for it. We use this input as a sample of the quality of the individual's work (``writing sample") in two ways. First, we add it to their profile, visible to all users of the batch, so that they can themselves determine that individual's writing skills. Second, we also evaluate it separately, using an external crowd, for comparison purposes during our results' analysis. Users have three minutes to complete the individual writing sample stage.

\begin{figure}
    \centering
   \includegraphics[width=0.5\textwidth]{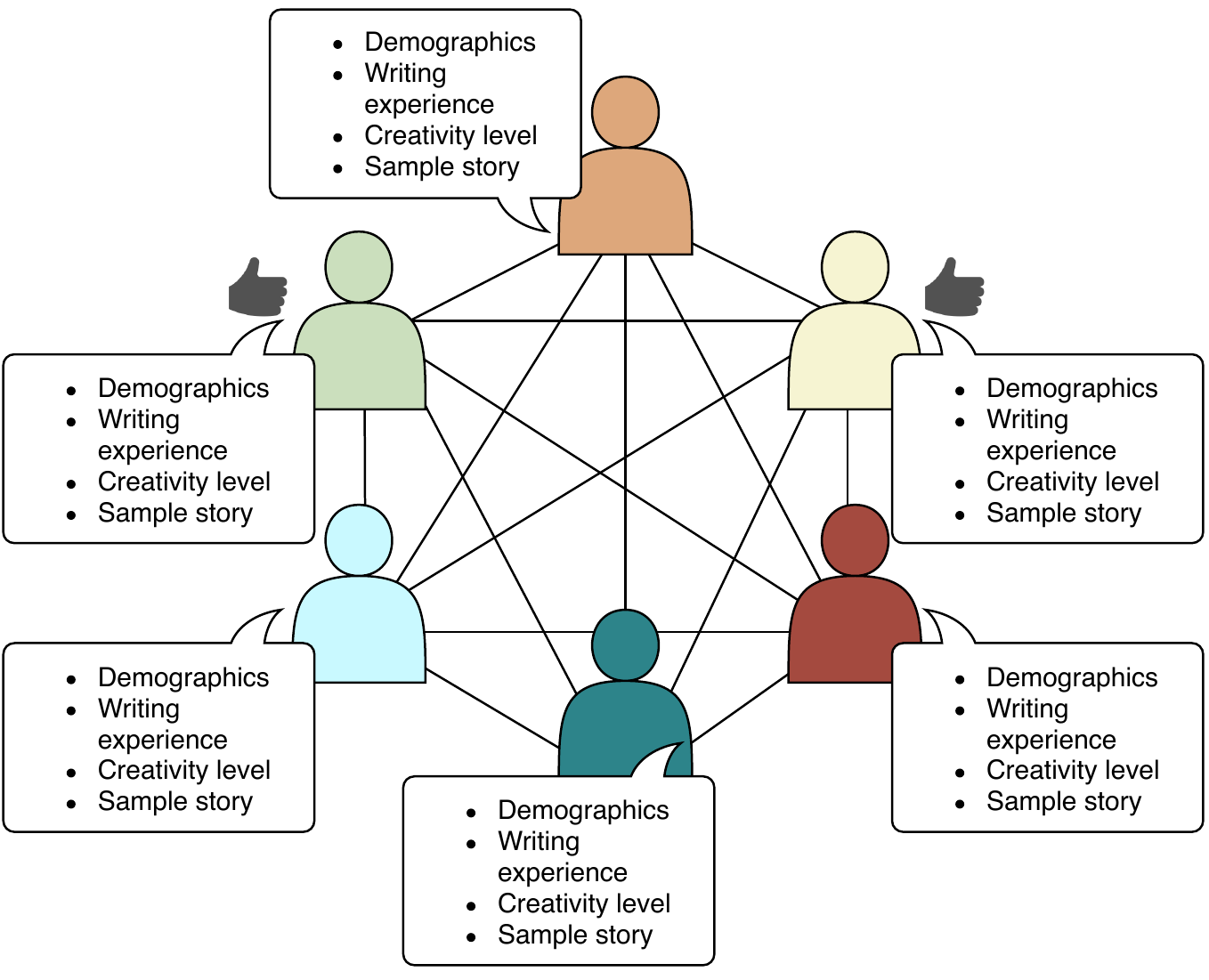}
    \caption{Self-organization -- Round 1. Self-organization takes place during the teammate selection phase. Participants vote for their preferred teammates based on their teammates' profiles (consisting of demographics, writing experience, creativity level and sample story), and the SOT algorithm uses these votes to form the teams. In the next rounds, participants also see the average rating of each person, as well as their own rating for that person (if that exists).}
    \label{fig:graph_voting}
\end{figure}

\subsubsection{Self-organization Decision: Teammate selection}\label{sec:teammate_selection}
Next, users are moved to the teammate selection step, illustrated in Fig.~\ref{fig:graph_voting}. Here they will select their preferred teammate(s) from the full list of user profiles of the batch. 
Users can see each others' profiles, where each profile contains the following information about them: (i) username, (ii) demographic information, and (iii) writing sample. They can also see the (iv) average rating each candidate teammate has received by his/her previous teammates (``others' rating") and (v) rating the person looking at the profile page may have given to that particular candidate teammate if they have already worked together in the past (i.e. ``own rating"). Note that items (iv) and (v) are only shown from the second round onward after users have already collaborated at least once.

In the teammate selection stage, users will be asked whether they want to work with the same teammate or not. Users are also asked to indicate up to two other candidate teammates to work with.  These latter choices are useful to the system for two reasons: i) in case the user indicated that they do want to work with their previous teammate, but that teammate is unavailable or ii) in case the user indicated that they no longer want to work with their previous teammate. 
The SOT algorithm (Section~\ref{sot_algorithm}) will use these choices to construct a ``preferences matrix'' and form the teams of the next round. 

The teammate selection stage is a critical step in self-organization. It demands users to quickly assess multiple sources of information across multiple users, and balance potentially conflicting candidate decisions: e.g. the psychological safety of working with a person similar to them ~\cite{mcpherson2001birds}, versus the choice of choosing a highly rated person with whom they might not have a lot in common. In the next rounds, when the information available to the users for making a choice increases, users will also need to individually assess their relative gain from continuing with the same teammate (lower communication overhead and presence of transactive memory ~\cite{hollingshead2003potential}, since the team has learned to work together) versus the risk of losing the chance to work with a new teammate (for example a previous round winner) who could potentially increase their chances of writing that round's winning story. Users are given two minutes to choose their preferred teammates.

\begin{figure}[t]
    \centering
  \includegraphics[width=1\textwidth]{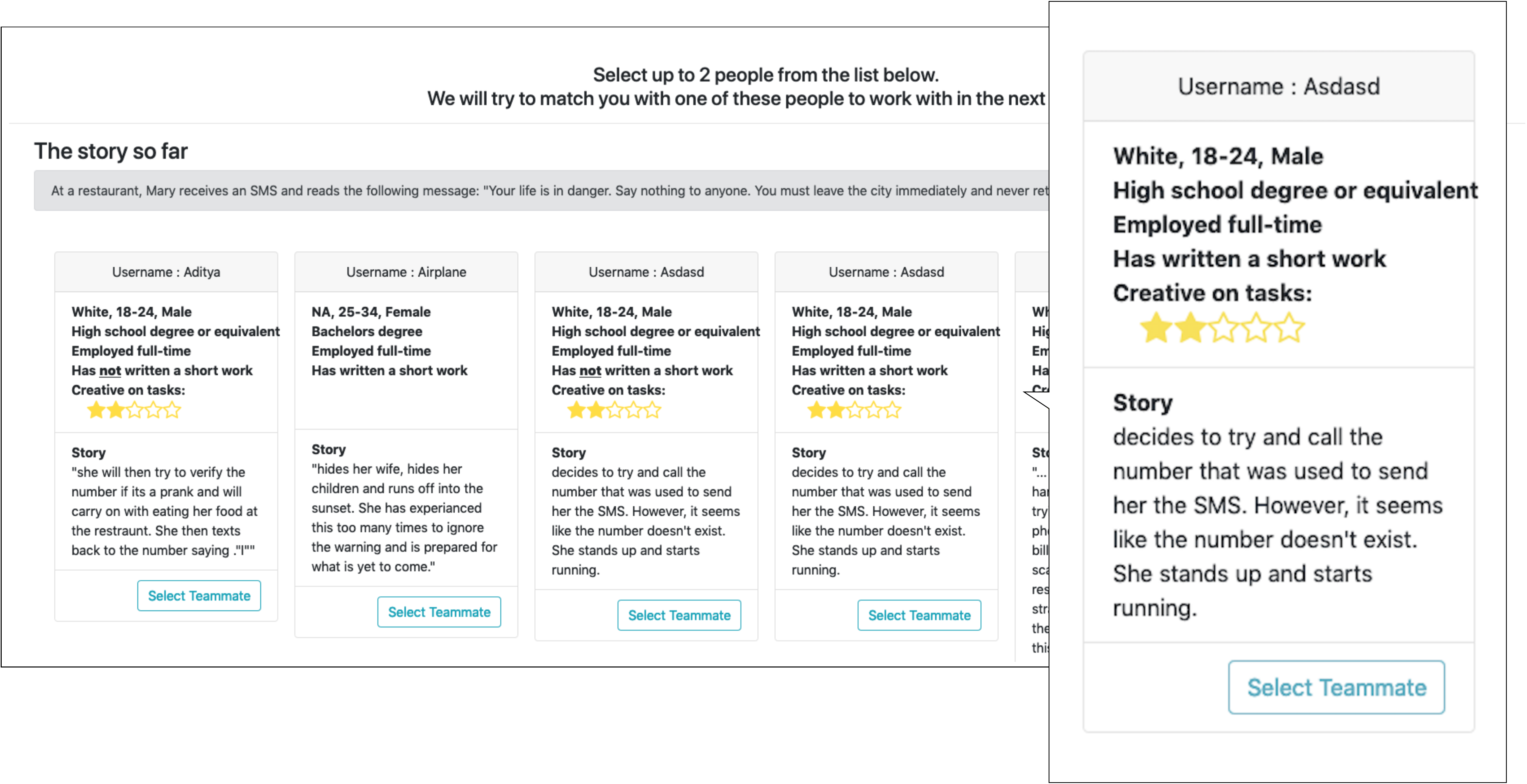}
    \caption{Participant profiles as seen by other users in our experimental platform interface -- Teammate selection phase, Round 1. In the next rounds, participants also have the choice to indicate if they wish to stay with their previous teammate or not.}
    \label{fig:profile_participants}
\end{figure}

\subsubsection{Collaboration: Teamwork and internal evaluation}\label{sec:team_collaboration}
As soon as the algorithm has placed users in teams based on their indicated preferences, each team is moved to an online synchronous collaboration space, with a text writing area and capability to chat\footnote{The software Etherpad (\url{https://etherpad.org/}) was used to facilitate the teams' synchronous collaboration in this study.}. Here each team is instructed to continue the story so far (``main story''). In the first round, the main story is simply the initial pre-authored story presented to the users at the individual writing sample stage. In the next round, the main story will gradually increase, since after every round the winning team's story will be appended to it. Team members are free to discuss and work together to continue the main story, in any way that they like. This allows us to observe different team dynamics and interaction patterns, work and collaboration strategies, creativity patterns, etc. Each collaboration round lasts for four minutes. Thirty seconds before time is up, users also see a reminder to wrap up their story.

Once time is up, the teams are asked to evaluate one another on three axes and on a Likert scale of 1 to 5. (i) Skillfulness (``How skillful was [teammate's username] in continuing the story?"), (ii) Collaboration ability (``How good is [teammate's username] as a collaborator?"), (iii) Helpfulness (``[teammate's username] comments were helpful"). Each team member is also asked to assess their own helpfulness level (``My ideas and comments were helpful") on a Likert scale from 1 to 5. Finally, users are asked to assess the number of core competencies they noticed having in common with their teammate (``[Teammate's username] and I were similar in:''), with four possible options (multiple or none can apply): (i) task commitment (``Commitment to working hard on this task''), (ii) work strategy (``How we think the work should be done''), (iii) Skill similarity on task (``General abilities to do a task like this''), and (iv) Personal values (``Personal values'').  These ratings are used to enrich the profile of each user, both in terms of the ``other's ratings'' (average rating by previous teammates) and in terms of the ``own ratings (of the person looking at that user's profile), as explained earlier (Section~\ref{sec:teammate_selection}). Team members have half minute to complete their evaluation of one another. 

We determined the timeline of the experiments after several experimental trials with multiple combinations of time slots. When adjusting these time slots we also took into consideration the AMT outsourcing model, which favors micro-tasks. Although extending the time for each phase of the task could have been beneficial to some workers, we noted that most were able to produce their judgment within the given time. Batch sizes did not differ greatly between experiments and the teams' stories that needed to be voted on by each worker were no more than three at a time and considerably short in length. With batches significantly larger than what we used in this study, lengthier time slots would have been even more so applicable. We discuss further the scalability of the system in Section \ref{discussion}. 

\subsubsection{Competition: Voting for Best Story and Presenting the Winning Team}\label{sec:vote_best_story}
After evaluating their teammate for that round, each individual user votes for their preferred story continuation, among the $S-1$ candidate continuations, $S$ being the total number of teams from the previous round (users cannot see or vote for their own team's continuation). In voting for the best story, users can see which team (i.e. which two usernames) produced which story continuation. Users have one and a half minute to read and decide on their preferred continuation. Once the time is up, the story with the most votes (``winning story'') is presented to them, along with the usernames of the two members of the winning team. The profile of winning team members is updated so that the bonus amount for winning is added to their individual total earned reward. Presenting the winning story before users are asked to make a decision on their teammate of the next round is important to give users an overview of their results so far. From a task point-of-view, the story peer assessment at this stage allows for a collective decision to emerge regarding the outcome of the task, i.e. users collectively have full control over the task result. 
Peer review is also a proven way of incorporating quality assurance during the task~\cite{whiting2017crowd}. Alternative ways of evaluating the team result after each collaboration round can be envisioned and they are straightforward to incorporate, without affecting the core of the proposed system. These ways include assessment by an external crowd or by one person, such as the client who commissioned the task. 

Next, and assuming the predetermined total number of rounds is not over, users return to the teammate selection stage. As explained in Section~\ref{sec:teammate_selection}, here they must decide whether they want to continue with the same teammate as in the previous round, or whether they want to change. In both cases, they are also asked to indicate up to two additional candidate teammates, from the full candidate teammate profile list, for the algorithm to use either in case it cannot accommodate their first choice (in case they wanted to stay with their previous teammate), or for the algorithm to use to match them with a suitable alternative teammate (in case they wanted to change). 

The cycle of self organization-collaboration-competition continues, with the main story gradually increasing in length as more and more team continuations are appended to it. After a number of rounds, which for the purposes of this study is set to three, users see the final story, the final user ranking (in a descending order based on the number of times a user has been a member of a winning team), and a final questionnaire about their overall experience. Once they fill in this questionnaire, users are redirected to the crowdsourcing platform and paid.

\subsection{Self-organization algorithm}\label{sot_algorithm}
The SOT algorithm is one of the first to maximize user agency in algorithm-based crowd team formation. Its aim is to assist but not dictate the self-coordination process, matching users with those teammates that they mostly prefer working with. The algorithm receives as input the individual user profile ratings, encoded in a tabular form as follows. Assume user $A$ has worked with user $C$ in the previous round. If a user $A$ indicates that he/she wants to continue working with $C$ in the next round, then $A \rightarrow C=3$. If user A indicates that he/she does not want to continue working with C, then $A \rightarrow C=0$. User $A$ can also select from the list of other candidate team members, with whom he/she has not worked in the previous round. Assume this list contains users $B$ and $D$, and assume the user indicates $B$ as a preferred teammate for the next round, and does not indicate $D$. Then $A \rightarrow B=2$ and $A \rightarrow D=1$. In brief, a user's preference is given the highest weight to those teammates that the user has worked with and wants to continue working with, and the lowest weight is given to those that the user has worked with but does not want to continue working with.

\begin{figure}
\centering
\subfloat[Bidirected graph, all profile ratings]{
        \includegraphics
        [height=2.7cm]
        {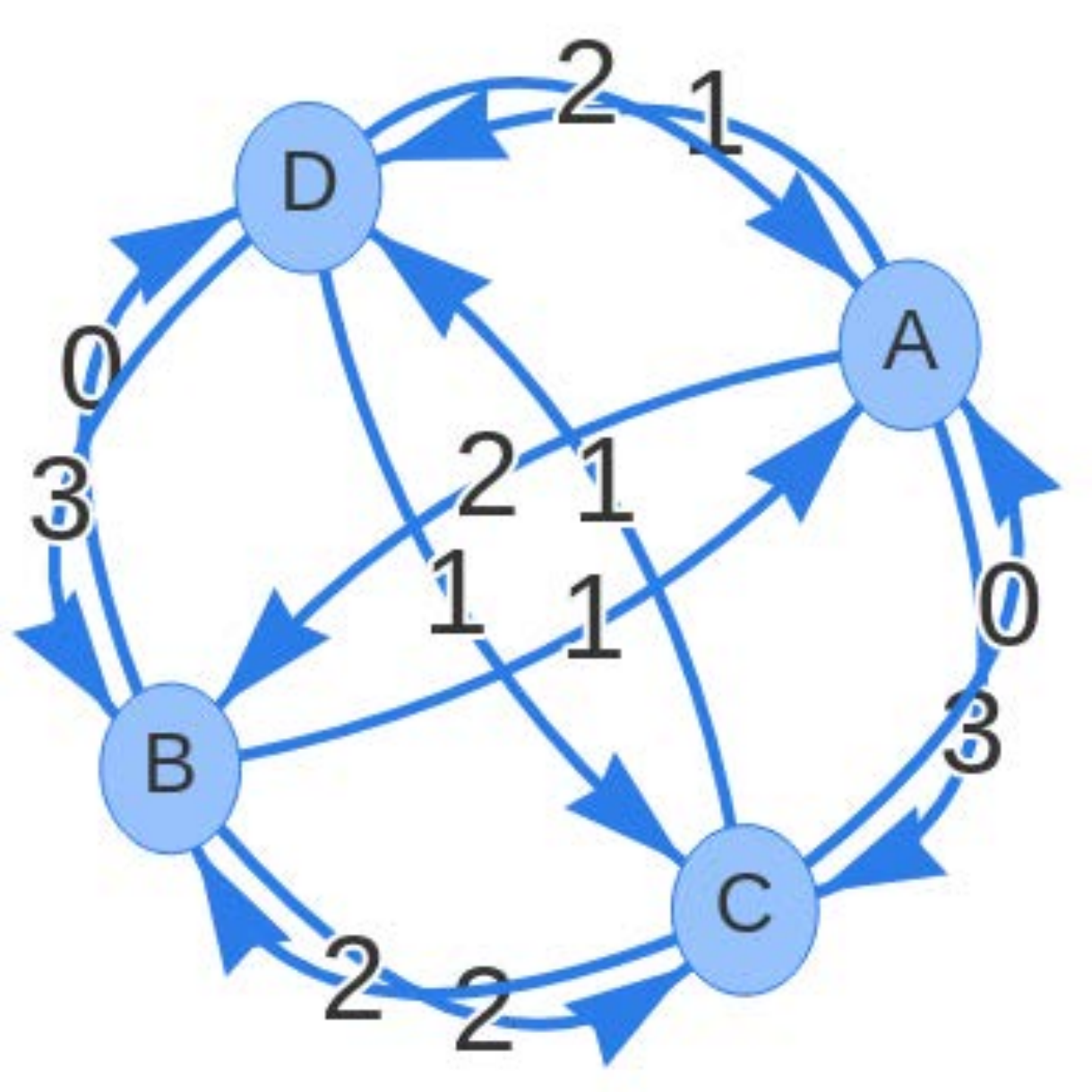}
        \label{fig:direct_graph}
}\hfill
\subfloat[Affinity graph]{
        \includegraphics
        [height=2.7cm]
        {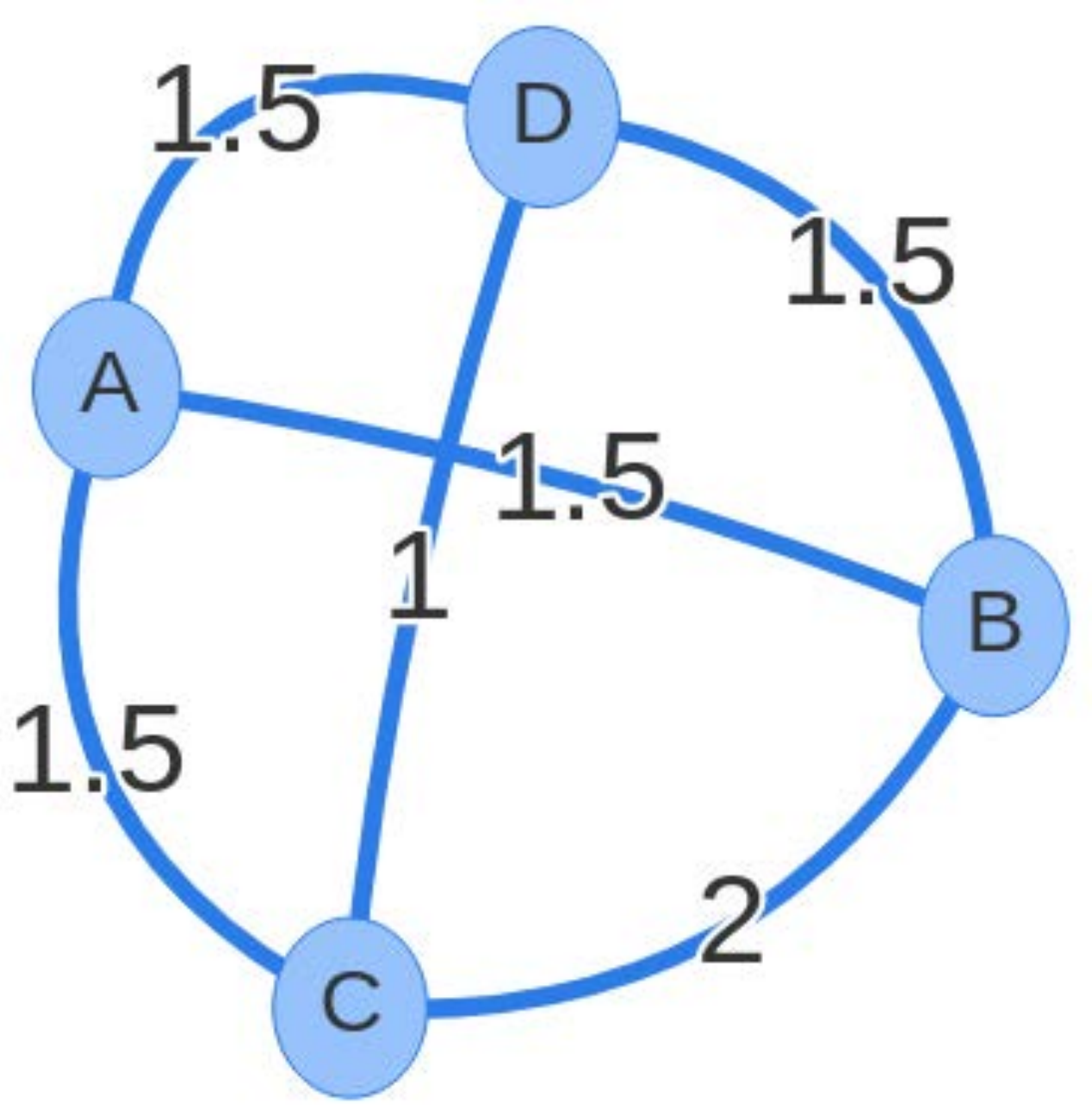}
        \label{fig:affinity_graph}
}\hfill
\subfloat[All possible affinity graph cuts]{
        \includegraphics
        [height=2.7cm]
        {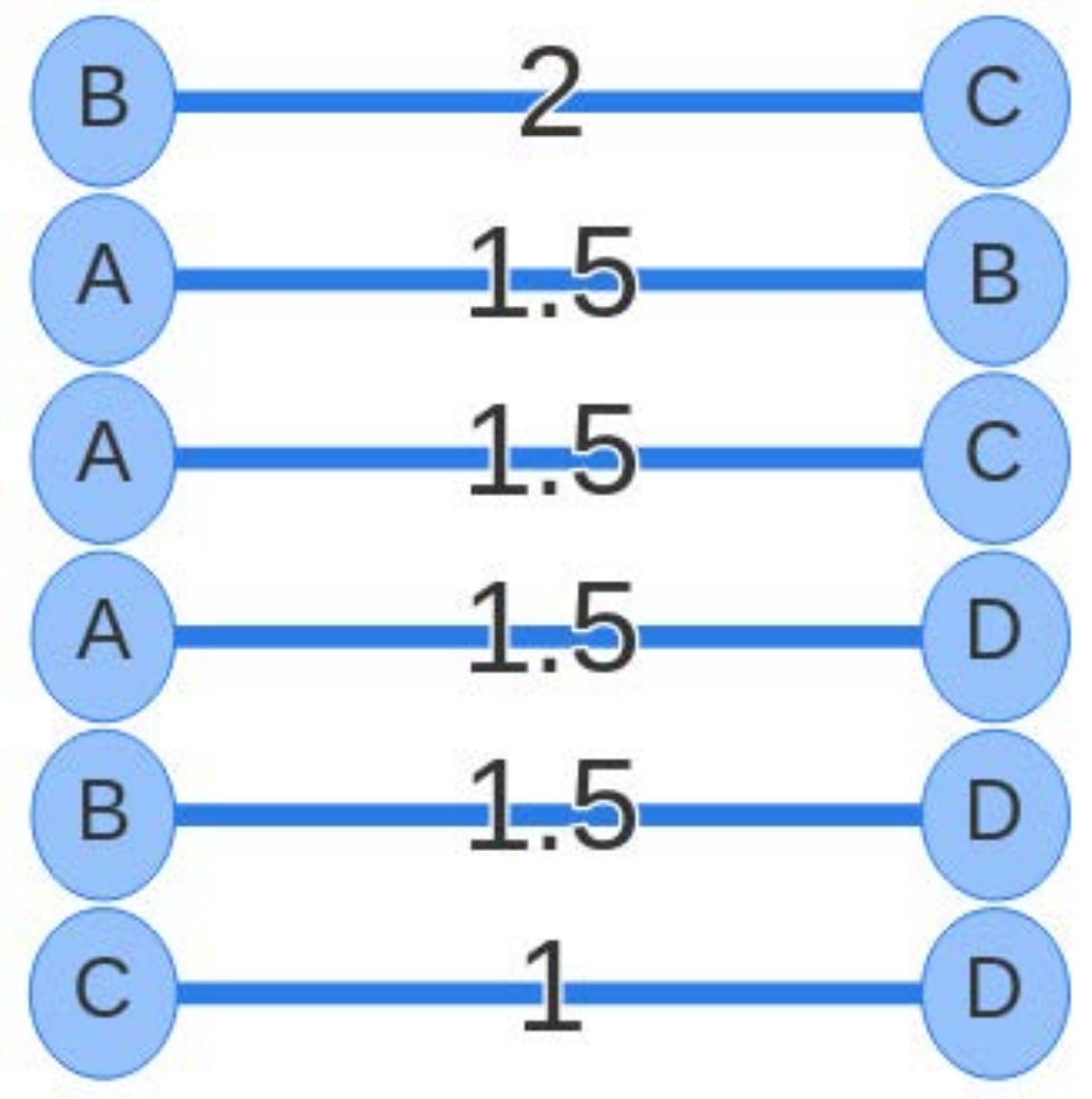}
        \label{fig:subsets_of_two}
}\hfill
\subfloat[Selected affinity graph cuts]{
        \includegraphics
        [height=2.7cm]
        {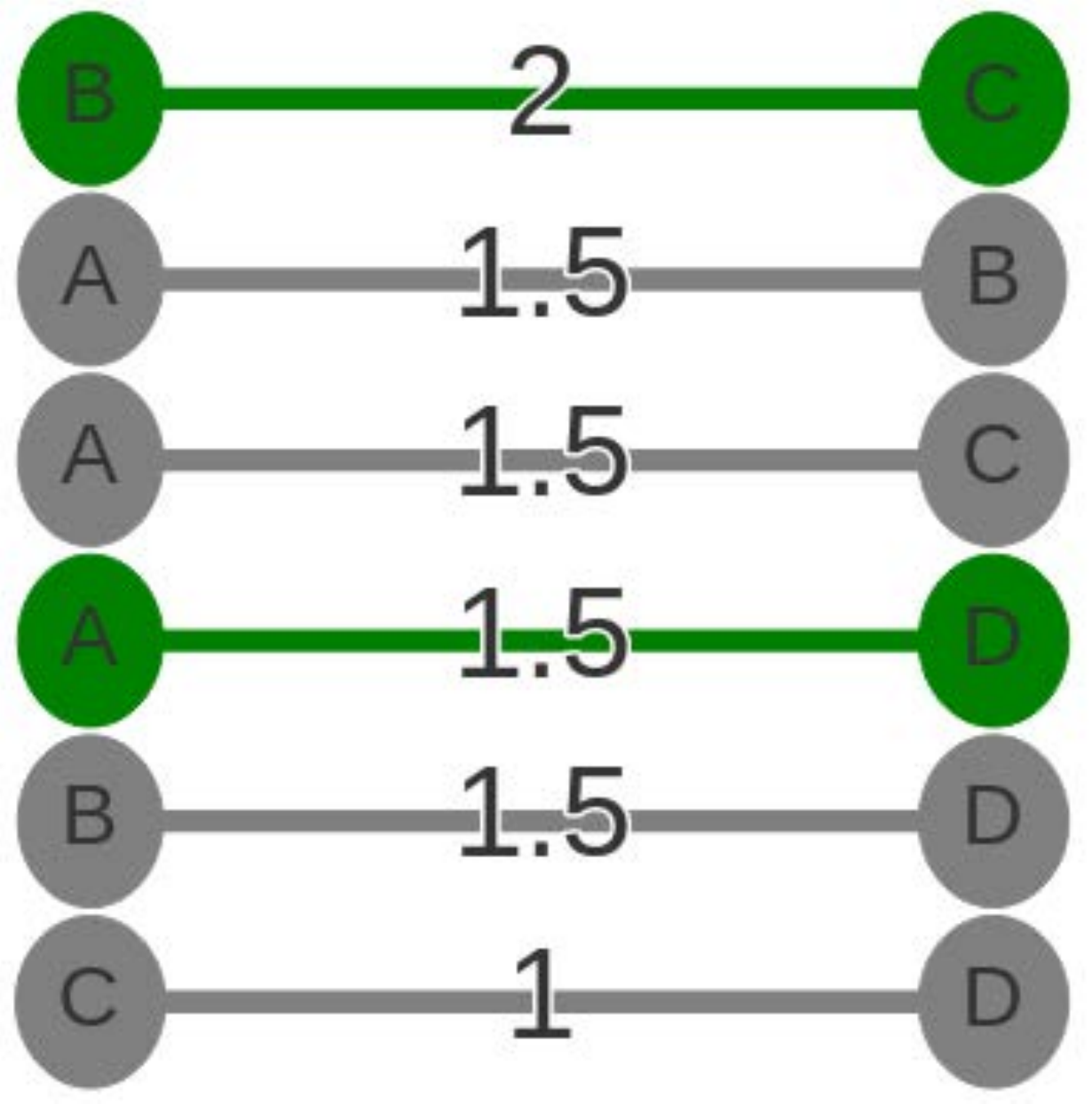}
        \label{fig:subsets_of_two_selection}
}
\caption{Steps of the algorithm's operation. Nodes represent the participants, edges represent participant teammate preferences (values 0-3, higher values mean higher teammate preference). The algorithm (a) first constructs a complete bidirected graph comprising all user preferences, and then (b) it constructs the affinity graph comprising the average pairwise ratings. Next (c) it ranks all possible teams (i.e. graph cuts) in descending order of collaboration preference and (d) respectively forms the teams.}
\label{fig:major}
\end{figure}

Using a user rating vector per user (i.e. two per team) as shown in Fig.~\ref{fig:direct_graph}, the algorithm then constructs a complete graph (``affinity graph") with candidate team members as nodes, and the  average pairwise ratings between individual users as the edges (Fig.~\ref{fig:affinity_graph}). Next, the algorithm identifies all possible candidate teams, i.e. all possible graph cuts of size two. Next, it ranks the candidate teams on a list based on their average pairwise rating (edge value) from the highest to the lowest (Fig.~\ref{fig:subsets_of_two}). In other words, the algorithm ranks the candidate teams starting from those that want to work together again, continuing with those that have not worked together before but would like to, and ending with those that do not want to work together again. 
From this ranked list, the algorithm selects the first team, and removes all other candidate teams that contain the selected team's members (as one person can only be in one team at a time). Fig.~\ref{fig:subsets_of_two_selection} shows the selected team of $B$ and $C$ in green and removes gray nodes containing users $B$ or $D$ as options. The algorithm continues in this manner, until the list of candidate teams is empty, and all users have been placed in a team. In case of ties, the algorithm chooses randomly. The pseudo-code for this process is shown in Algorithm~\ref{alg:sots-optimal}.

\begin{algorithm}
 \KwData{Individual profile ratings}
 \KwResult{Final list $F$ of teams for next round}

 Create complete graph $G=(V, E)$: $V$: candidate team members, $E$: average pairwise ratings\;
 Find all possible graph cuts of size 2 (candidate teams), $\rightarrow C$\;
 Sort $C$ in descending order\; 

 $F \gets \emptyset$\; \label{alg:emptyS}
  \While{$C \neq \emptyset $}{
  Pick first  element $C_i$ = \textless x,y\textgreater{ }in $C$: x,y $\in V$\;
  $F = F \cup \{C_i\}$\;
  $C = C  - C_i -\{C_j\}$: x $ \in C_j \mid $ y $\in C_j$\;
 }
 \Return{$F$}\; 
 \caption{Self-organizing team formation algorithm. The algorithm creates the teams based on the ratings and preferences of the users regarding their candidate teammates. 
 }
\label{alg:sots-optimal} 
\end{algorithm}

Next, we will describe the experimental conditions we designed to study our methods, which shed light on how team self-organization affects quality of work, team satisfaction and collaboration styles.

\subsection{Experimental Conditions}
For this study we work with three experimental conditions, one examining the proposed approach and two benchmark conditions. 

\begin{itemize}
\item \textbf{SOT}: This condition stands for ``Self Organizing Teams'', and studies the proposed approach of self-organization in online teams. People are given the choice to indicate their preferred teammates, including the option to stay with the same teammate that they worked with in the previous round. The algorithm respects these choices and aims to place each person with the teammate of their choice.

\item \textbf{Placebo}: This is the first benchmark condition. This condition creates the illusion of agency, where participants believe that they can self-organise, but eventually they cannot do so. The workflow and experimental interface of this condition is identical to that of SOT, i.e. users do have access to the teammate selection stage. The difference between the two conditions happens in the background, where participant choices at the teammate selection phase are \textit{not} taken into account by the algorithm. Instead, each participant is paired with a randomly allocated teammate, and stay with them for the entire task.

\item \textbf{No-Agency}: This is the second benchmark condition. It is made to resemble existing methods of placing people in ad-hoc online teams, where users lack agency over teammate selection. It is also similar to the benchmarks used in the literature for testing online team formation from a worker crowd (see for example~\cite{Salehi:2017:HCS:2998181.2998300}).
In this condition, participants are paired randomly with one teammate at the beginning of the task, and stay with them throughout the process, without the option to choose their teammates (neither functional nor placebo). The difference with the Placebo condition is that the task workflow of the No-Agency condition skips the teammate selection stage entirely. 

\end{itemize}

\subsection{Participants}
\label{sec:participants}
A total of 140 people took part in a total of 18 experiments for this study. Eight of these dropped out due to internet connection issues, resulting in a final total of 132 people who finished the experiment. The study participants were recruited either as university students or as Amazon Mechanical Turk workers, in batches of 4 to 12 people depending on availability. 
The batches of participants belonging to these two different user groups (paid and volunteered) were managed in separate sessions and they were equally distributed between the conditions.
The allocation of people to condition was made in a round robin manner to avoid biases due to participant type or batch size, resulting in six batches per condition.

The total number of people who participated in the Placebo condition was 52 and those participating in the SOT condition was 48, and those participating in the No-Agency condition was 32. 

To further exclude the possibility of confounding factors, we conducted a series of post-hoc checks. First, using the demographics information filled in by the participants in the beginning of the experiment (Section~\ref{fig:profile_participants}), the sample was controlled for statistically significant differences across the conditions in terms of demographics, namely gender, age, ethnicity, education, employment status, prior experience and self-perceived creativity. An Analysis of Variance (ANOVA) showed no significant differences across any of the aforementioned axes (all at $p>.1$).
A similar analysis also excluded any statistically significant differences in terms of individual writing skills between the three conditions, as evaluated by the external crowd evaluators who rated each participant's individual writing sample, again in the beginning of the experiment, on the axes of grammar and syntax, interest, originality, plot structure and overall impression of the story sample ($p>0.4$ across all evaluation axes). 
Finally, we controlled on whether there was any difference in user perceptions of the benefit of agency across the three conditions. An analysis of variance comparing perceived teammate selection usefulness, from the final questionnaire answers, showed no statistically significant difference across the conditions ($p>.6$).
These are the same criteria that, as we will see later, were used to evaluate the stories produced by the teams during the collaboration. Having a sufficiently balanced sample across the three conditions, we proceed with the analysis of our results.   

\section{Results}\label{sec:results}
We organize our results as follows. First, we look into the quality of the produced work by different teams in the three conditions to investigate the question, ``Did the teams formed under the SOT condition produce stories of higher quality than those of the Placebo and No-Agency benchmark conditions?''.
Second, we look into the quality of the collaboration to investigate the question, ``Did participating in the SOT condition enable participants to collaborate better and be more satisfied with the process of collaboration, compared to participants in the Placebo and No-Agency benchmark conditions?''.
After answering these key questions we look deeper into the mechanics of self-organization, examining two emergent patterns of self-organization, namely the presence of an objective function that drives the participant collective, as well as network clustering phenomena.

\subsection{Work outcome quality: SOT teams write stories of higher quality}
A total of 196 unique story continuations were produced by the teams. The final winning stories were 18. To evaluate the quality of these stories, we employed a crowd of external judges, hired through AMT. Each story continuation was evaluated by 10 AMT workers, on a ten-point Likert scale (1-10), and on five quality criteria: grammar and syntax (``How grammatically and syntactically correct is the story?'', ranging from ``Not correct'' to ``Very correct''), interest (``How interesting is the story?'', ranging from ``Not interesting'' to ``Very interesting''), originality (``How original is the story?'', ranging from ``Not Original'' to ``Very Original''), plot structure (``How good is the story plot?'', ranging from ``It doesn't make sense'' to ``It flows nicely''), and overall impression (``Overall how much did you like the story?'', ranging from ``Not at all'' to ``Very much''). These criteria were selected as they are among the most frequently used by professional short story evaluators~\cite{boden2004creative}, and because they represent a balanced mix of both objective (grammar, plot structure) and subjective (interest, originality, overall impression) axes~\cite{diaz2015evaluating}. 

\begin{figure}[t]
    \centering
   \includegraphics[width=0.8\textwidth]{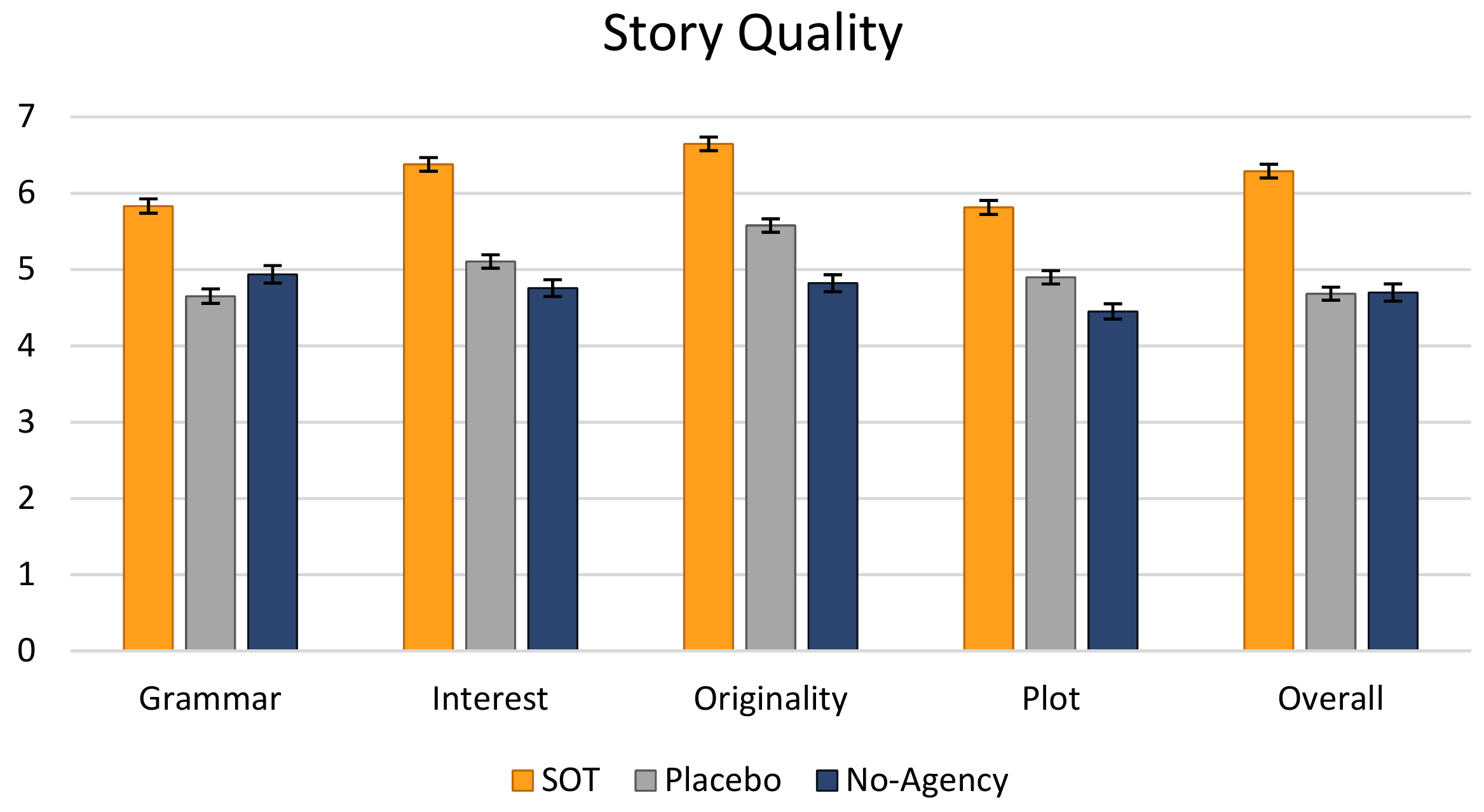}
    \caption{SOT teams produced stories of higher quality compared to the teams of the two benchmark conditions as rated by external evaluators. Results across all five axes are statistically significant at $p<.001$}. 
    \label{fig:story_quality}
\end{figure}

 An analysis of variance indicated that SOT teams create stories of significantly higher quality than the benchmark condition teams, across all five quality criteria, albeit with slightly different absolute value differences between the conditions (Figure~\ref{fig:story_quality}). Specifically, the SOT team stories were rated higher in terms of grammar ($m_{grammar}=5.83, SE_{grammar}=.094$), interest ($m_{interest}=6.38, SE_{interest}=.09$), originality ($m_{original}=6.65, SE_{original}=.089$), plot structure ($m_{plot}=5.81, SE_{plot}=.092$) and overall impression ($m_{overall}= 6.29, SE_{overall}=.091$), compared to the respective evaluations received by the Placebo stories ($m_{grammar}=4.65, SE_{grammar}=.099$, $m_{interest}=5.11, SE_{interest}=.088$, $m_{originality}=5.58, SE_{originality}=.089$, $m_{plot}=4.9, SE_{plot}=.087$ and $m_{overall}=4.68, SE_{overall}=.085$), as well as those received by the No-Agency stories ($m_{grammar}=4.94, SE_{grammar}=.114$, $m_{interest}=4.76, SE_{interest}=.11$, $m_{originality}=4.83, SE_{originality}=.111$, $m_{plot}=4.46, SE_{plot}=.109$ and $m_{overall}=4.7, SE_{overall}=.113$), with 
$F_{grammar}(2,1957)=41.835, p<.001, \eta^{2}=0.063$,
$F_{interest}(2,1957)=78.742, p<.001, \eta^{2}=0.095$, $F_{original}(2,1957)=84.198, p<.001, \eta^{2}=0.101$, $F_{plot}(2,1957)=50.391, p<.001, \eta^{2}=0.075$ and $F_{overall}(2,1957)=99.847, p<.001, \eta^{2}=0.103$. A Tukey post hoc test per quality axis also revealed that SOT groups differed significantly from the other two benchmark conditions across the five quality axes (at $p<.001$), while the Placebo and No-Agency conditions differed significantly in terms of interest ($p<.05$), originality ($p<.001$) and plot ($p<.05$). 
A regression analysis analyzing the story continuation data, with round as a random effect, showed that the round does not account for the relationship between the higher performance of the SOT condition compared to the others. 

Finally, a Pearson correlation analysis between ratings for attribute vs. the other (like grammar and overall) showed that the rater's evaluations were significantly correlated (p < 0.01) with each other, indicating the presence of the ``halo effect'', which is well-documented in many social judgment settings. The Halo effect implies that a rater's judgments of one quality dimension  tends to influence others, even in the presence of sufficient information to allow for an independent assessment of them~\cite{nisbett1977halo, saal1980rating, woehr1994rater}. See Table~\ref{table:pearson_external_eval} in the Appendix for detailed values.

Overall, the external ratings on multiple factors show that stories produced in the SOT condition were better quality compared to the benchmark conditions. Next, we will discuss the perceived quality of their collaboration by participants in different conditions.

\subsection{Collaboration quality}
One of the major goals of team formation is for individuals in the team to collaborate effectively. While collaboration effectiveness can be studied using many methods, we first focus on the perceived collaboration by individuals in the team.

 \begin{figure} 
    \centering
   \includegraphics[width=0.8\textwidth]{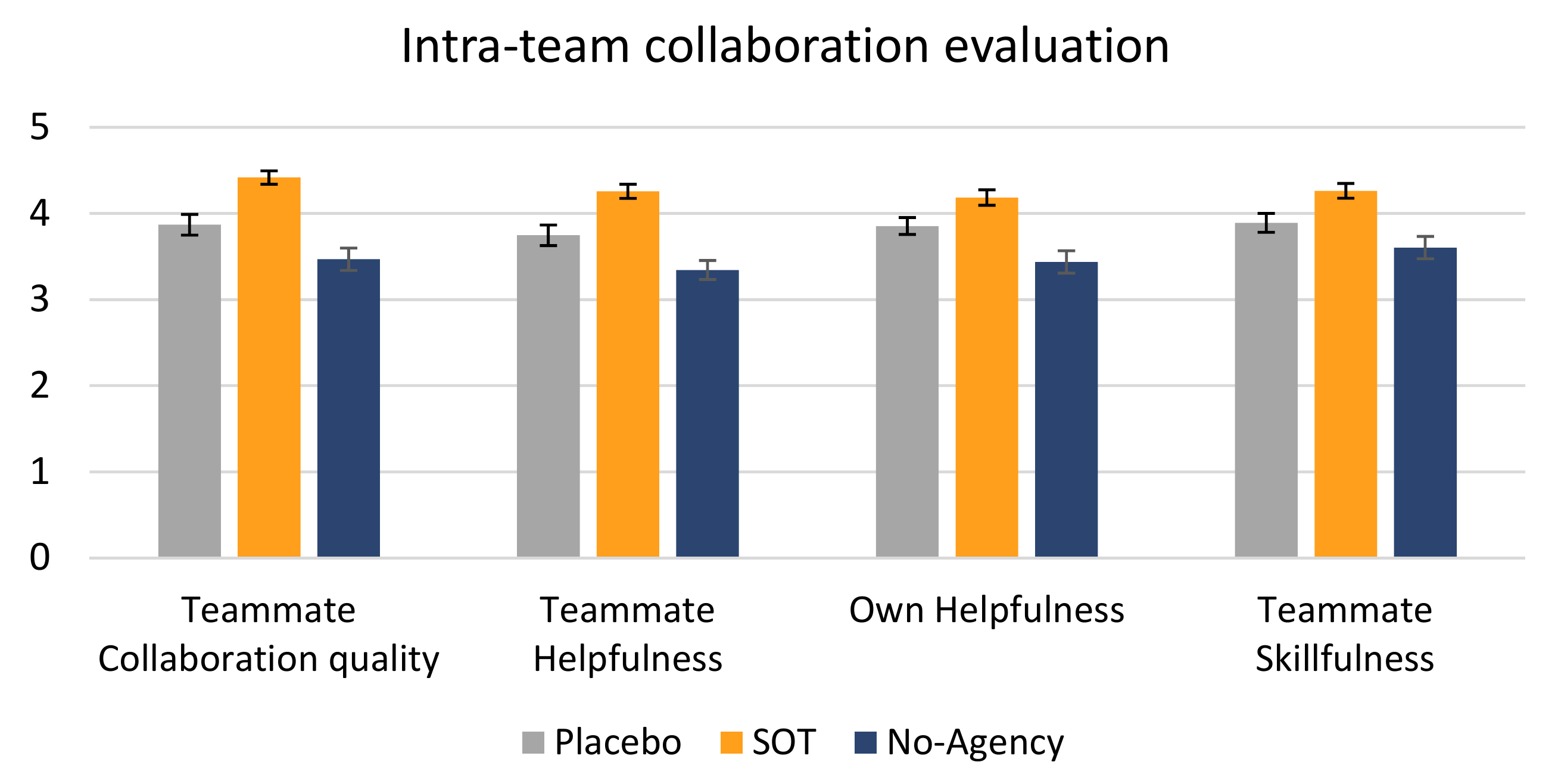}
    \caption{SOT team members rated their teammates higher in terms of collaboration quality, helpfulness, skillfulness, and perceived their own contributions as more helpful to the team's final output, compared to both benchmark condition teams (all four axes at statistical significance level $p<.001$).}
    \label{fig:intra-team_collaboration}
\end{figure}

\subsubsection{People in SOT condition evaluate each other higher in collaboration, helpfulness and skillfulness}\label{sec:peer_evaluations}

On average, the team members in the SOT condition rated each other significantly higher as collaborators (on a scale of 1 to 5)
($m_{collab}=4.42, SE_{collab}=.078$) and in terms of how helpful they were ($m_{help}=4.26, SE_{help}=.082$), compared to teams formed under the Placebo condition ($m_{collab}=3.87, SE_{collab}=.12$, $m_{help}= 3.75, SE_{help}=.12$), as well as compared to teams formed under the No-Agency benchmark condition ($m_{collab}=3.47, SE_{collab}=.13$, $m_{help}= 3.34, SE_{help}=.11$). 
These results are confirmed by one-way ANOVA analyses $F(2,211)=21.364, p<.001, \eta^{2}=.213$ and $F(2,211)=8.089, p<.001, \eta^{2}=.209$, for the metrics of collaboration ability and helpfulness levels respectively. For each of these results, a Tukey post hoc test was also run, revealing the presence of statistically significant differences between the SOT condition and each of the benchmark conditions, as well as between the two benchmark conditions. Specifically, for the metric of helpfulness, the groups formed under the SOT condition evaluated one another significantly higher compared to the groups formed under the other two benchmark conditions ($p<.001$ for each comparison), while there was also a statistically significant difference between the Placebo and No-Agency benchmark conditions (with $p<0.05$). For the metric of collaboration ability, the groups formed under the SOT condition also evaluated one another significantly higher compared to each of the groups formed under the other two conditions ($p<.001$ for each comparison). Here too, the post hoc test revealed that the Placebo groups perceived their teammates as more collaborative than the groups formed under the No-Agency condition ($p<0.05$). 

We find that the perception of helpfulness within the collaboration went both ways. Not only SOT team members perceived their teammate's contribution as more helpful, but through the collaboration they also perceived their own contributions to the team as significantly more helpful ($m_{ownHelp}=4.19, SE=.09$). 
In contrast, participants in both the two benchmark conditions found that their ideas and contributions were not as helpful for their team ($m_{ownHelp}=3.86, SE=.098$ for the Placebo condition and $m_{ownHelp}=3.60, SE=.13$ for the No-Agency condition), with $F(2,211)=8.089, p<0.001, \eta^{2}=.119$. A Tukey post hoc test showed that the aforementioned results were only significant as to the difference of the SOT groups with the benchmark groups (with $p<0.05$ between the SOT and Placebo condition and $p<0.001$ between the SOT and No-Agency condition). Interestingly however, there was no statistically significant difference between the Placebo and No-Agency groups ($p=0.262$) in terms of how they perceived their own helpfulness, although as we saw before, the two groups did differ in how they perceived the helpfulness of their teammate (No-Agency was lower).

We also observe that SOT team members perceived their teammates as significantly more skillful ($m_{skill}=4.26, SE_{skill}=.086$) compared to the perception that the team members have of their teammates' skills for both benchmark conditions ($m_{skill}=3.89, SE_{skill}=.11$ for the Placebo condition and $m_{skill}=3.44, SE_{skill}=.13$ for the No-Agency condition), with $F(2,211)=15.366, p<0.001, \eta^{2}=.177$. A Tukey post hoc test revealed that these differences are significant among all three groups, with $p<0.05$ between the SOT and Placebo conditions, $p<0.001$ between the SOT and No-Agency condition and $p<0.05$ between the Placebo and No-Agency condition.
Interestingly the aforementioned higher perception of skillfulness is not because SOT members are indeed more skillful; in fact, as also mentioned in Section~\ref{sec:participants}, participants in the three conditions do not differ statistically in terms of skillfulness as evaluated by external evaluators on their individual writing samples. Previous research ~\cite{hansen2002impact} demonstrates that when people are more satisfied by their collaboration, then they tend to think more highly of their peer, thus being more prone to associate affective trust to positive expectation about belonging to that team. For similar reasons, group cohesiveness can positively affect the perception of satisfaction and team performance.

We note that from the three conditions, the No-Agency benchmark condition, i.e. the one where people were not given any (not even a placebo) option to choose their teammate was the one with the lowest intra-team evaluations in terms of all four axes of collaboration, helpfulness and skillfulness.

These results, summarized in Fig.~\ref{fig:intra-team_collaboration}, indicate that the teams formed under the SOT condition are more satisfied during their collaboration, and able to collaborate and help each other more, despite not being objectively more skillful than the individuals of the benchmark condition teams.

\subsubsection{SOT team members are more aligned as to how the work should be done}
\label{sec:items_in_common}
During their peer evaluations the participants of each team also indicated which, if any, competencies they had in common with their teammate. As explained in detail in Section~\ref{sec:team_collaboration}, they could report a common sense of commitment to the task, work strategy similarity, skill similarity, and/or common personal values.

\begin{figure}
    \centering
   \includegraphics[width=0.8\textwidth]{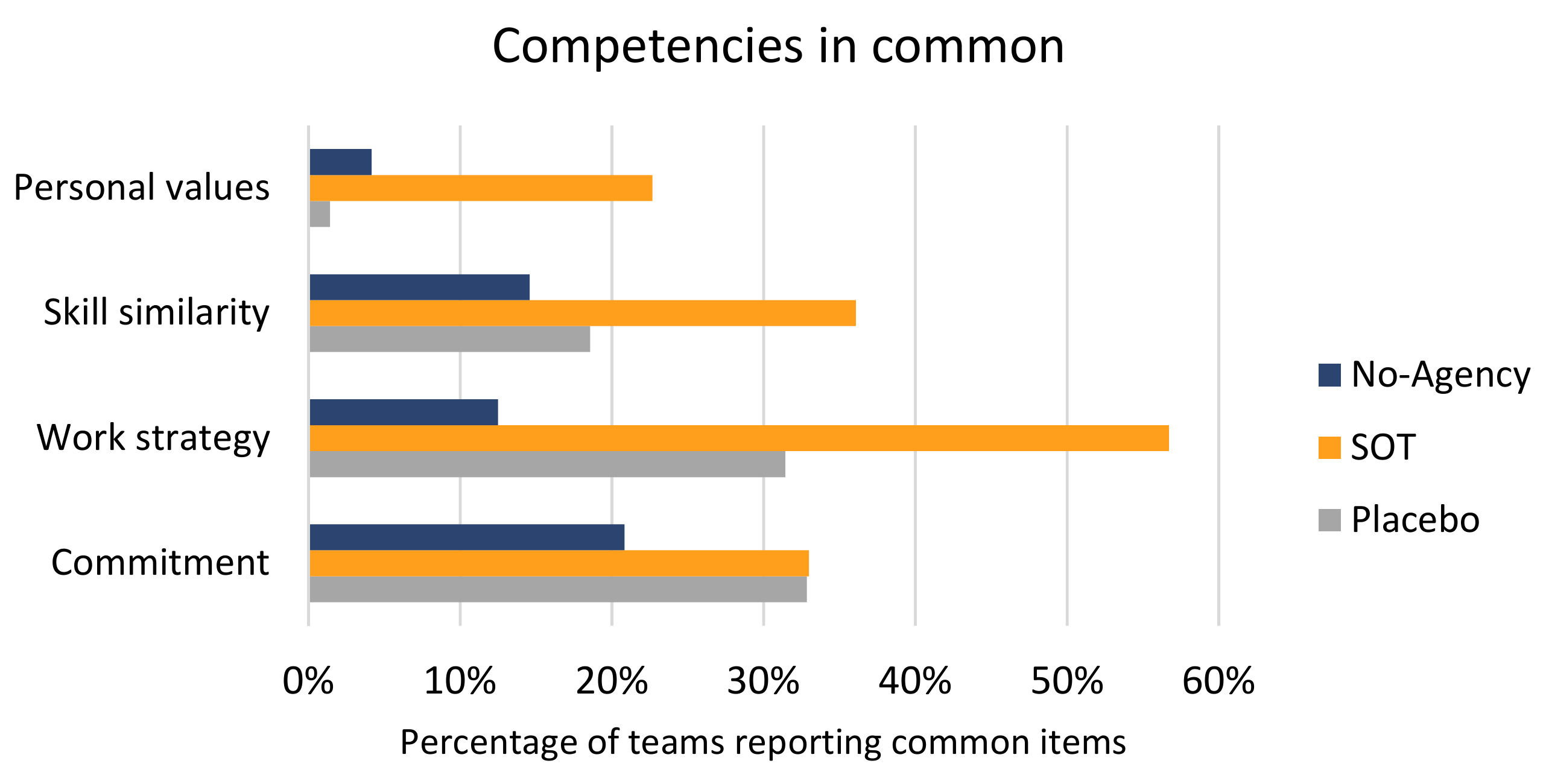}
    \caption{Percentage of teams reporting common competencies across the three conditions. SOT team members reported being significantly more similar in terms of work style, at $p<.001$, compared to the other two benchmark conditions. The teams did not differ significantly in their perceived skill similarity, personal values or commitment to the task.}
    \label{fig:workflow}
\end{figure}

A chi-square test of independence was performed to examine the relation between condition and number of work style items that the teams reported having in common. The relation between these variables was significant, $\chi^2(8, 215)=32.29, p<.001$, with SOT team members reporting to have more items in common with their teammate compared to both benchmark condition teams. Post-hoc tests also showed that the only item where the teams differed significantly was on how similar they reported to be in terms of their work strategy (``how we think the task should be done'') ($\chi^2(2, 215)=17.76, p<.001$), while they did not differ significantly in their perceived skill similarity, personal values or commitment to the task. In other words, the only factor found to distinguish the SOT teams from the benchmark ones was their high similarity on how they thought they should approach the task.

This result indicates that when people's choice of a teammate (SOT condition) is honored, then they tend to pair with teammates with whom they share common work practices, confirming prior literature~\cite{tenney2009being}.

\subsubsection{SOT teams write shorter stories and have more turn-taking equality}
Teams assembled under the SOT condition produced shorter story texts ($m_{textpad}=234$ characters, $SE=13$), compared to both the Placebo ($m_{textpad}=320$ characters, $SE=16$), and the No-Agency condition teams ($m_{textpad}=556$ characters, $SE=88$), with $F(2,212)=17.395, p<0.001, \eta^{2}=.885$. 

\begin{figure}[t]
    \centering
   \includegraphics[width=1\textwidth]{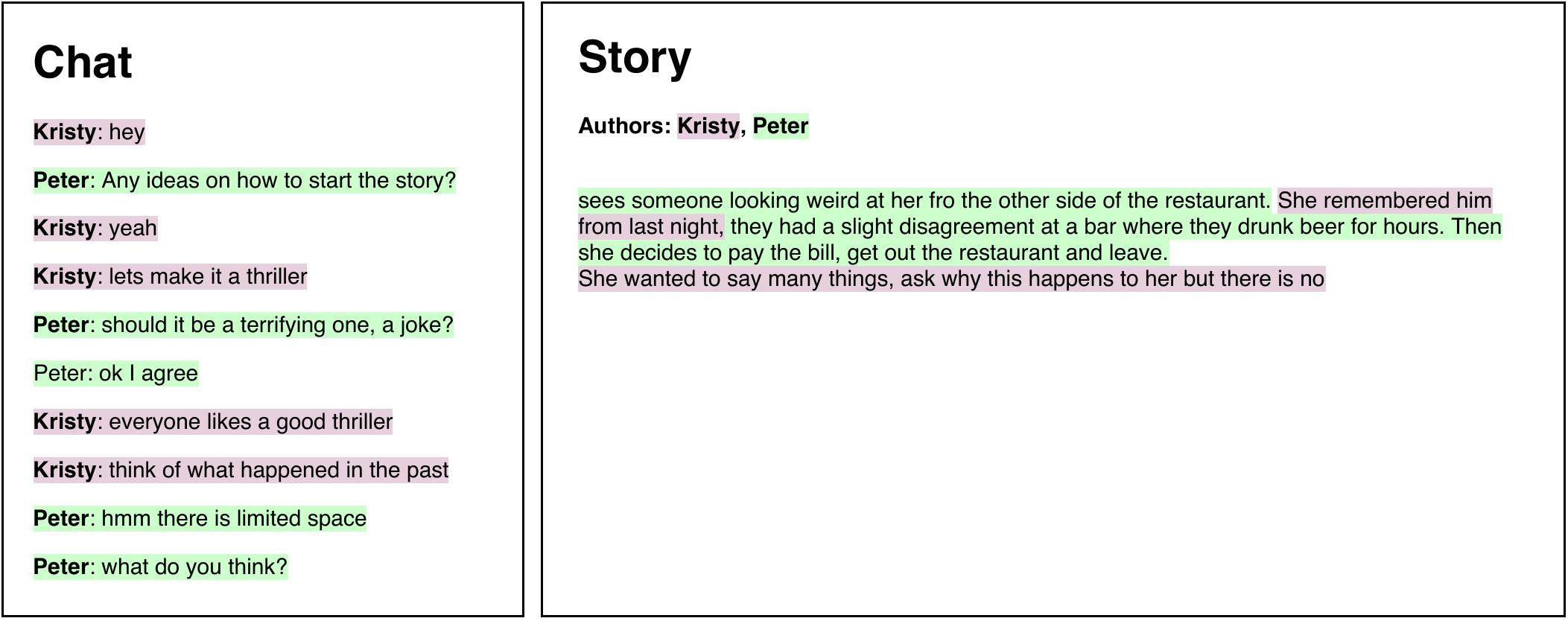}
    \caption{Overview of a collaboration space and chat under the SOT condition }
    \label{fig:optimal_story}
\end{figure}

A Tukey post hoc test revealed that these differences were significant among all three conditions, with $p<0.05$. One explanation for this could be that the benchmark condition teammates know each other better, since they have worked together in the past rounds, whereas the same is not always true for SOT teammates, who may or may not have worked together in the past. If that is the case, then the chats between the three conditions could be expected to differ statistically in length, with the SOT teams chatting more in an effort to establish a common ground in each round; therefore having less time to work on the actual task. However, the analysis of variance comparing the total chat length between the three conditions showed that there is no statistically significant difference. 

Looking deeper into the process of story writing, we examine the way that the teams in the three conditions produce their common story text by measuring a metric for ``turn-taking''. 
Turn-taking is a property of collaboration~\cite{sacks1978simplest} based on construction contribution which allows two or more entities to build a discourse from separate units. The metric was chosen for the evaluation of the given experiments as it considers the amount and the timing of the individual contribution towards the group work.   

\begin{figure}[t]
    \centering
   \includegraphics[width=1\textwidth]{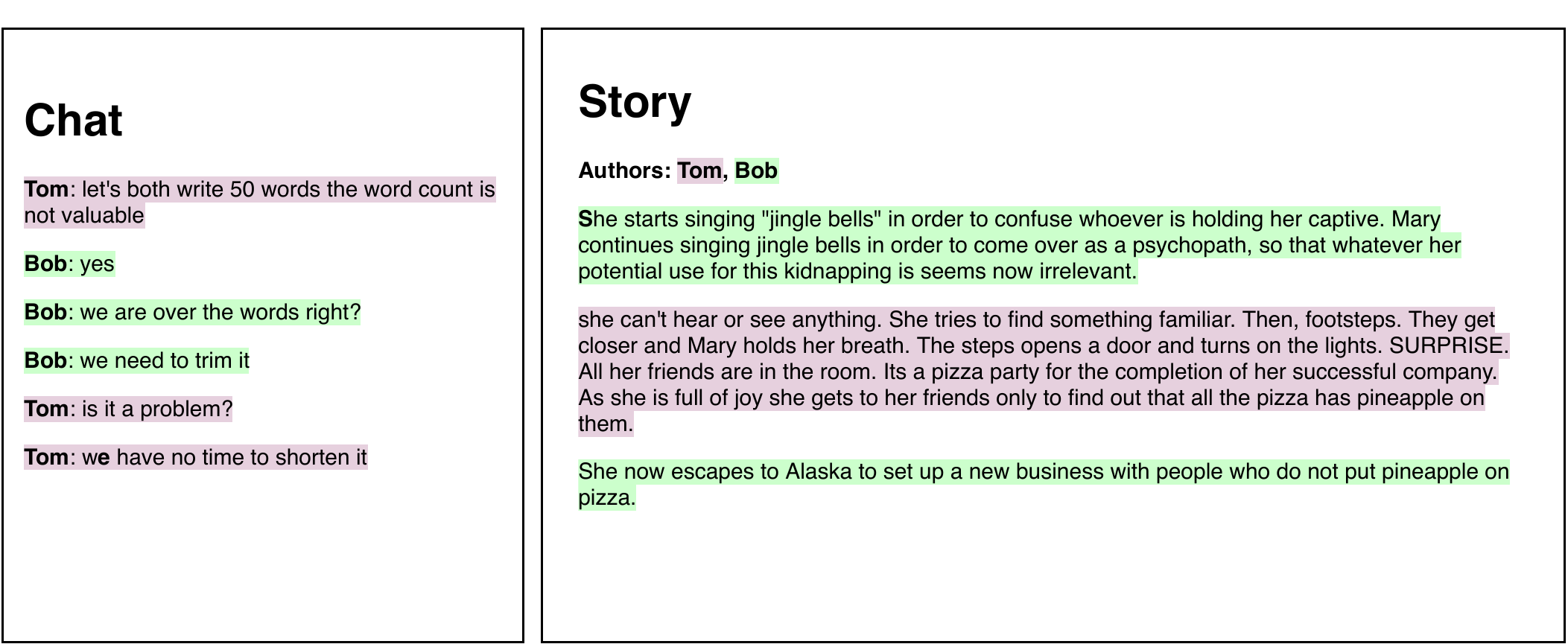}
    \caption{Overview of a collaboration space between two teammates under the Placebo condition }
    \label{fig:placebo_story1}
\end{figure}

We measure turn-taking as follows. First we identify every text piece (every segment of text entered by the users) each team member entered in the common text area, and the order in which this member entered this piece. This gives a sequenced order of contributions. For example, assume a team consisting of person \textit{A} and person \textit{B} and a writing sequence of their collaboration to be \textit{\{ABABAAA\}}. We encode this sequence as \textit{\{-1,1,-2,2,-3,-4,-5\}}, sum and normalise it by the sequence length. 
The turn-taking was tracked by the back-end part of the system, which saved the final version of each story continuation when the writing phase ran out of time. The software used for hosting this kind of synchronous collaboration, called Etherpad \cite{erdal2017using}, helped to automatically highlight the text in a different color for each user, meaning that user A could see his or her text in a different color than that of User B. Figures \ref{fig:optimal_story} and  \ref{fig:placebo_story1} illustrate the above.

Every time the contribution of one team member is followed (``matched") by the contribution of the other team member, the value of this metric is equal to zero. The more person A dominates the writing process, the more negative values the metric receives. The more person B dominates the writing process, the more positive the metric becomes. Hence, values around zero indicate a balanced writing process in terms of turn-taking.

The analysis of the story writing processes for the two conditions, using a random allocation of team members to position \textit{A} and \textit{B} of the metric, shows that the SOT teams have significantly more balanced text logs, in terms of turn-taking style ($m_{turn}=0.01, SE=.051$) compared to the Placebo ($m_{turn}=-0.22, SE=.064$) and No-Agency condition teams ($m_{turn}=-0.50, SE=.091$), with $F(2,212)=14.515, p<0.001, \eta^{2}=0.315$. A  Tukey post hoc test revealed that these mean differences were significant among all three conditions, at $p<0.05$.

This result, combined with the fact that SOT team members are more satisfied with their collaboration (Section~\ref{sec:peer_evaluations}) and have more common working styles (Section~\ref{sec:items_in_common}) than the benchmark condition teams, indicates that overall SOT teams seem to collaborate more harmoniously in writing their common stories.

Next, we will show two sample stories, one from the SOT and one from a benchmark (Placebo) condition, to qualitatively show different type of chat logs and stories. One story is a result of a productive collaboration, while the other shows how teams may not communicate well, leading to below par stories. Stories and chats from the No-Agency condition follow a similar pattern as those from the Placebo condition, as we also saw in the quantitative analysis above.

\paragraph{Sample Stories.}
Figure~\ref{fig:optimal_story} shows an example chat and collaboration workspace between two people, Kristy and Peter, in the SOT condition. In line with the statistical results discussed above for SOT teams, we observe that
the two teammates collaborate on the story creation by taking turns in both the chat and the story workspaces. The continuation of the sample story is made of equal chunks of text written and discussed by the ad-hoc pair of workers, who discuss about the online collaboration through a fairly balanced chat thread and the use of open questions. Workers seem to work synchronously, and build on each other's contribution in a manner that coherently continues one another's previous statements.

Figure~\ref{fig:placebo_story1} shows an example chat and collaboration workspace between two people, named Tom and Bob, in the benchmark Placebo condition.
We observe that Tom and Bob agree to split the workload into separate chunks and then they work individually: each essentially writing a different continuation of the preceding winning story. This strategy of collaboration did not produce the best optimal outcome since the resulting plot was incoherent and unbalanced. Also worthwhile noting is that the first paragraph written by Tom  "{\fontfamily{qcr}\selectfont She starts singing 'Jingle bells'} .. " got moved to the top of the story just a few seconds before the end of round, whereas Bob's contribution got abruptly pushed underneath it. While this example is not representative of all chat logs and stories in the Benchmark condition, it does illustrate the lack of collaboration and quality that was also confirmed by the statistical results discussed above for that condition.

\subsection{Looking deeper into the mechanics of self-organization}
After examining how self-organization affects the work and collaboration quality of participating teams across the conditions, we now take a deeper look into how this method affects more subtle behavioral elements of team formation.

\subsubsection{Strategic voting}
So far, we have seen that SOT teams tend to collaborate better, feel more satisfied by their collaboration, and produce better work results. We now look into what motivates people to team up the way they do. For this, we compare the SOT and Placebo conditions, since these are the only ones where people were given the option to select their teammate (although this option is not honored in the Placebo condition). Since the participants of the Placebo condition do not eventually change teams, we look at \textit{voting intention}, as revealed during the teammate selection stage of each condition. In this stage, which takes place after each collaboration session as explained in detail in Section~\ref{sec:teammate_selection}, participants get to indicate which teammate they would like to work with in the next round. We first examine whether people tend to select previous winners as their preferred teammates. Indeed, an Analysis of Variance shows that, regardless of condition, the winners of previous rounds gather significantly more profile votes on average 
($m_{win}= 5.89, SE=0.377$) compared to individuals who have never won in the past ($m_{non-win}= 4.04, SE=0.377$), with $F(1,98)=11.431, p=0.001, \eta^{2}=0.171$.

\subsubsection{Winners choose teammates based on winning potential, non-winners choose those whose profile they like the most}
The aforementioned strategic voting strategy seems to pay off. In the question ``What mattered the most when choosing a teammate'' 
at the end of the task, we observe that winning participants were twice as likely to report that they chose the person that would make them win (28\% of winners, i.e. 22 out of 79) compared to non-winning participants (8 out of 55, i.e. 15\% of non-winners). In contrast to winners, who seem to choose strategically their teammate, non-winners were twice as likely to choose people whose profile information they liked the most (11\% of non-winners versus only 4\% of winners). 

\begin{figure}
\centering
\includegraphics[width=0.8\textwidth]{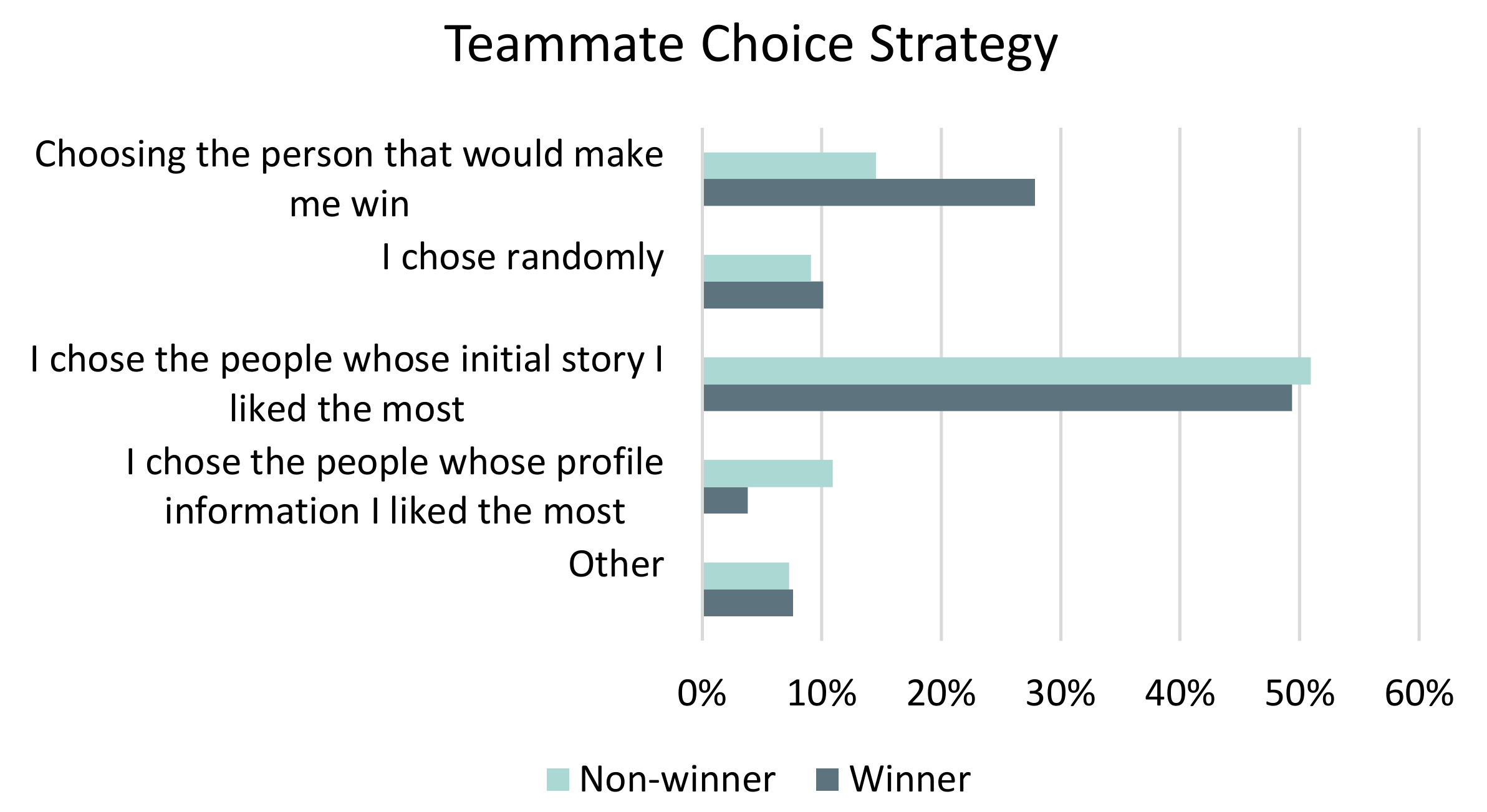}
\caption{Participants self-reported strategies of selecting a teammate. Winners, i.e. people that had been in a winning team at least once, were twice as likely to purposefully select teammates that would help them win. Non-winners were twice as likely to choose people whose profile information they liked the most. The rest of the answers received similar percentages by both winners and non-winners. }
\label{fig:teammate_choice_strategy}
\end{figure}

These two elements, i.e. winners driven more by a ``playing to win'' strategy and non-winners driven more by their teammate's profile information, were the only answer items that distinguished winners and non-winners, as the rest of the participants' reported answers to this question received relatively equal percentages of answers (Fig~\ref{fig:teammate_choice_strategy}). We also note that from the other teammate selection strategies, choosing based on skill, i.e. a prospective teammate's individual writing sample, was the one preferred by most participants (almost 50\% both in winners and non-winners alike), but as we saw it did not, in the end, make a real difference as to winning probability.

\subsubsection{Teams who change often are less likely to win} Of the teams that belonged to the SOT condition, 26\% (10 out of 39) stayed together across 3 rounds, 28\% (11 out of 39) stayed together for two rounds and the remaining 46\% (18 out of 39) of the teams worked together only once across the three rounds. The results show that most of the users changed their teammates often (at least two times) whilst only a minority of those (26\%) stayed together for the entire duration of the task. Teams that stayed together for 3 rounds won 55\% of the times (10 victories across the total 18 rounds of the SOT experiments), those that stayed together two times won 28\% of the times (5 victories across the total 18 rounds) and those that stayed together only one round won 17\% of the times (3 victories across the total 18 rounds). Even when considering only their own cohort (comparing teams that stayed together across the same number of rounds), we see that the teams who stayed together three times had a 30\% chance of winning (10 victories out of the possible 30). Those teams that stayed together for two rounds had a 22\% chance of winning (5 victories out of 22). For those teams that stayed together for one round only,they had a 17\% chance of winning (3 victories out of 18). 
From the above we see that when considering users by their winning potential, those who won more are also those that decided to remain in the same team across all the given rounds.  

\subsubsection{Network analysis: SOT participants tend to form more distinct clusters of compatible teammates}
Finally, in a parallelism with machine learning and data-driven self-organization, we look for emergent patterns in the way that people ``cluster" across the three rounds, in the two conditions of SOT and Placebo (the option to self-organisation does not apply in the No-Agency condition by default). To do so, we represent the experimental batches as bidirectional affinity graphs, with each affinity graph consisting of a set of users (nodes) and a set of dyadic ties (edges) among user pairs, as explained in section~\ref{sot_algorithm}. Each edge between a pair of user nodes receives a value, which corresponds to their ``pairwise affinity", i.e. the intent of these two people to collaborate with or avoid one another, as denoted by their voting preferences at the end of each collaboration round.   
We then apply social network analysis, which provides a set of methods for observing the emerging patterns of the teams.

In our analysis of the graphs, we take into account the process of change produced by the dynamic connectivity of the vertices, with every worker accumulating votes across the rounds. The set of \textit{V} of vertices of the graph is fixed, whereas the set of \textit{E} edges changes with time, in an incremental fashion.  
The calculation of the final graphs is the same as the sequential analysis technique used for cumulative sum. 
The networks formed by the partial sums of weights describe the overall interactions between actors across the phases of the experiments. Where a pair of users consistently voted to remain together, the ties between the two nodes shorten, indicating stronger attraction for collaboration. Greater distances between nodes can be formed by repulsion as either one or both the teammates down-voted the other.   

The results from a preliminary analysis of the network topology show that the teams that worked under the SOT condition create on average more clusters or chains, unlike the benchmark condition, which display for the great majority of the graphs, dyadic clustering. Both in-degree and out-degree weights are considered. 
Stronger ties between workers creating larger clusters mean the potential creation of channels for interpersonal communication, determined by the person's choice of one or multiple teammate/s. More so, social networks formed under the SOT condition display stronger polarised attraction-repulsion mechanisms that are less detectable in the alternative condition. 

Considered together, the above indicate \textit{a pattern in human collaboration behavior that is similar to machine learning-based self-organisation: when given the agency to form their own virtual teams, and an ``objective function'' to maximize (which in this case is winning a reward), people do tend to actively explore their candidate teammate space and to gradually form clusters of ``compatible'' sub-groups that are fairly distinct and separate one from the other.} As we also explain in the Discussion section, further studies can be made in this very interesting direction, to examine whether other patterns observed in machine learning can also be observed in online work settings, such as the gradual stabilization of the compatible self-formed team clusters, how many exploration rounds this would require etc.

\begin{figure}
  \centering
  {\includegraphics[width=0.4\textwidth]{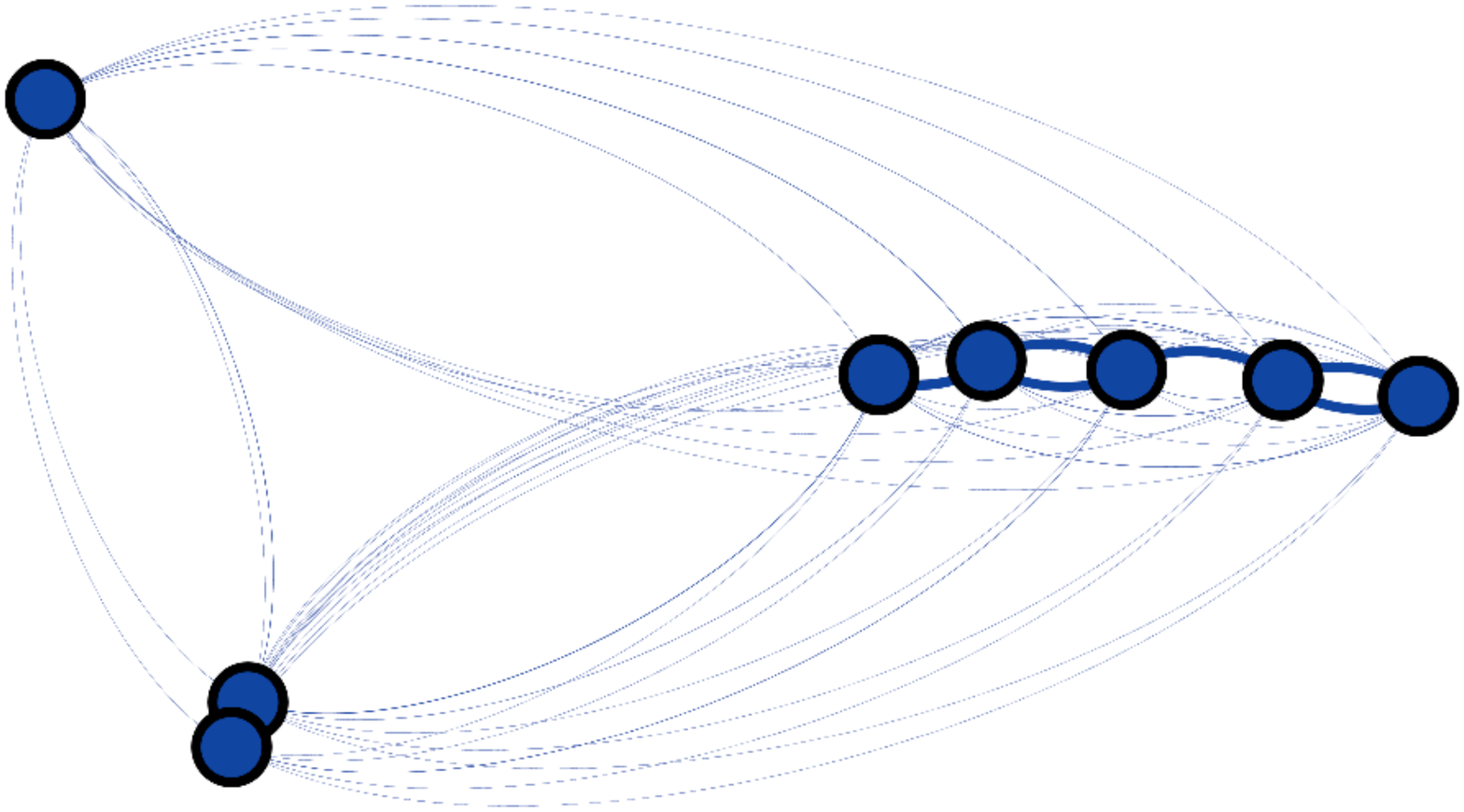}\label{fig:network_clustering_1}}
  \hfill
   {\includegraphics[width=0.4\textwidth]{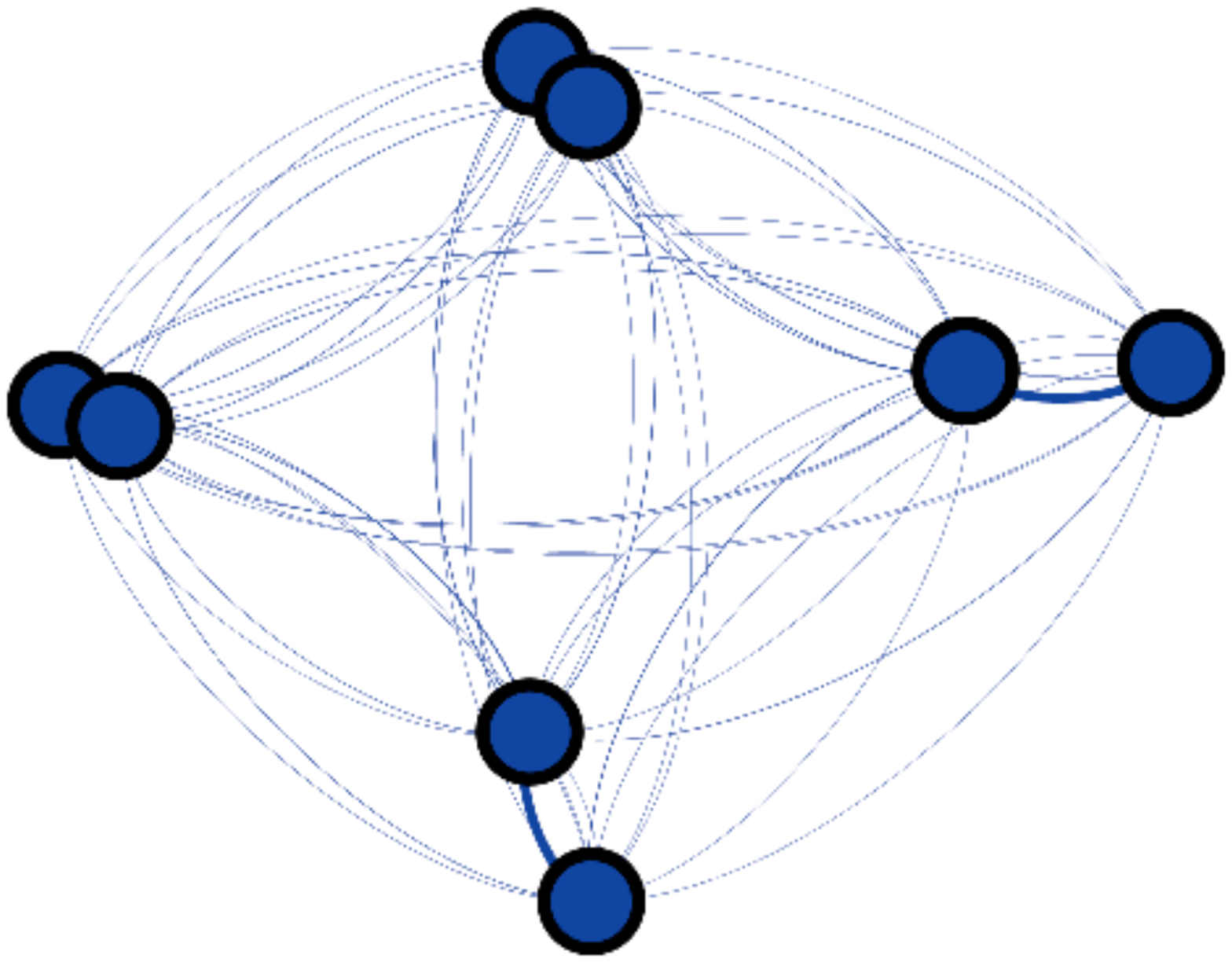}\label{fig:network_clustering_2}}
  \caption{Network clustering visualisation of the team formation process. Each node is a worker, and each edge is the team formation preference between that worker and a candidate teammate, as denoted by the voting matrices. The closer two nodes are, the more these two workers want to work together. In the SOT condition (left), we notice that participants form a large cluster, where a set of participants tend to prefer each other, while a few are not-preferred by most. In the benchmark Placebo condition (right), we do not observe a single large cluster.}
  \label{fig:network_clustering}
\end{figure}

\section{Discussion, Limitations and Future Work} \label{discussion}
In this work, we make a first attempt to explore the phenomenon of self-organization in the context of online work. Our results show that enabling participants to make their choice of teammate, and honoring this choice through an algorithm that aims to maximize intra-group preference, yields results of higher quality and teams that are more satisfied by their collaboration. A number of points merit further exploration and can be the object of future work. We discuss these points below.

\subsection{Objective function and task setting} 
The experimental setup of this study was set on a collaborative-competitive setting, which actively encouraged people to team up in such a way that their team would write the best story. The setting encouraged people to do so by significantly increasing their score or monetary reward every time their team's story would be voted as the winning one of the round. In view of the above, people chose strategically whom they would pair up with, showing an explicit preference for previous winners. Looking at this result from a macroscopic point-of-view, the worker collective adapted their objective function (i.e. how they make their teaming decisions) to their environment. This is in direct alignment with machine learning self-organization (e.g. self-organizing maps), where the objective function is determined by Euclidean distance, and the learning goal is to group the data points into clusters that minimize intra-group variance. The difference is that in the case of machine learning the objective function is known and given apriori by the algorithm designer, while in the case of human self-organization it can be driven (but not explicitly given) by external factors, such as the reward structure.

Observing the emergence of a collective behavior pattern, in our case the objective function, is also in line with prior work by Woolley and colleagues, who have consistently drawn attention to the presence of a general collective intelligence factor to explain group performance~\cite{woolley2010evidence, woolley2015collective} across participant contexts and cultures~\cite{engel2015collective}.

In the future, it would be useful to explore how to explicitly affect the objective function of the SOT by changing the environmental context of the task. For example, one could alter the reward structure, and instead of compensating participants higher for winning, to compensate them equally or with a minimal bonus, regardless of whether their team won the round or not. This could be expected to alter the way that people select teammates, potentially encouraging them to explore more diverse ways of teaming up and leading to different final work outputs (e.g. stories with different originality levels, plot structures etc.).  

Another factor that can affect the objective function in a self-organization context is the timing of specific prompts given to the participants. In our study, participants saw who were the winners of the previous round directly before choosing their teammates of the next round. This may be linked with the fact that, as we saw in section~\ref{sec:results}, many people intentionally tried to pair up with prior winners, in an effort to boost their chances of winning the next round. Changing this order, and asking participants to first indicate their preferred teammates before showing them the round's winner could have an effect on the way they form their teammate decisions, and would thus be interesting to explore as part of future work.

Although the most successful strategy, which in the end determined the objective function of the collective, was ``playing to win'', this does not necessarily means that all participants share the same goal. As we also saw in Figure~\ref{fig:teammate_choice_strategy}, the objective function of an individual may differ from that of the others within a batch. 

\subsection{Exploration-exploitation trade-offs} 
During the teammate selection stage, when making the decision of whether to switch to a new teammate or to stay with their previous one, participants are trading between the strategies of exploration and exploitation respectively. For the participants, opting for exploration means that they have an increased chance of finding the best, for them, teammate in the pool. However, exploring various pairings comes at the cost of a higher collaboration learning curve and more cognitively demanding interactions with the new teammate(s). The benefit of the exploitation strategy is that participants will work with a person they have already worked with, missing however the chance of discovering a potentially better teammate. These trade offs between exploration and exploitation have a direct impact on the performance of the team formation algorithm, and consequently on the performance of the batch as a collective.

First, exploration-exploitation is affected by the behavior patterns of the participants of the particular batch. The teammate decisions taken by the participants during each teammate selection round, shape the affinity matrix that the SOT algorithm uses to form the teams. For example, if most users do not opt for exploitation, meaning that they consistently vote to work with the same teammates, the affinity matrix is more sparse, providing less options for pairing the persons. One way to fix this would be to utilise the network analysis information, such as the one illustrated in Fig.~\ref{fig:network_clustering}, to potentially pair participants based on the clusters that are being formed, i.e. recommending ``teammates of teammates''. In the future it would be useful to explore whether such an approach, which is similar to collaborative filtering used in recommender systems, could help address such cold-start problems, while still ensuring user agency in teammate selection. An overall risk-averse batch of workers can also make the team formation algorithm susceptible to local optima: if people stick to their initial teammates, they will not have the chance to work with teammates with whom they might collaborate even better. To amend this, future work could explore allowing a certain degree of randomness in the SOT algorithm to enable some users to experience working with others, and therefore explicitly favor exploration. In recommender systems, this is achieved through the element of serendipity, and future work could explore introducing it at different degrees and time points.

Second, the exploration and exploitation patterns are affected by the choices of the task designer, for example in terms of the number of rounds and the batch size. In our study we used three rounds, and most participants reported that this number was sufficient, irrespective of condition\footnote{In the question ``The game had three rounds. Ideally, how many rounds would you need to decide who is the best teammate for you?'' at the questionnaire at the end of the task, most participants reported a value close to 3, with no statistically significant difference between the two examined conditions.}. Nevertheless, we noted that exploration had not finished at the end of the third round, since participants were still changing their teammate selection preferences. In the future, it would be interesting to examine how many rounds it takes for an average batch to converge, i.e. how many rounds it would take before the SOT collective stops exploring and starts to fully exploiting its stabilised team formations. Simulations could be a useful tool to explore this part of future work, as adding more rounds means lengthier and more costly tasks. Similarly, simulations could explore how the batch size affects convergence and the performance of the SOT algorithm. 

\subsection{Scaling, timing and worker cognitive capacity}
Attention is one of the most scarce resources of our information overloaded era \cite{davenport2001attention}. Although the estimates for the length of the human attention, in particular for general sustained tasks, have not yet been proven \cite{wilson2007attention}, a number of studies have pointed out that the type of task is the most influential factor in the quality and length of the average adult attention span. This is very much task-dependent, and, as supported by findings \cite{david2009essential}, it can even be conditioned by the engagement levels and intrinsic motivation of the individual. Attention, according to the attention economy, comes at a cost \cite{davenport2001attention}, and when intertwined with multiple concurrent activities, it becomes even more in short supply. In our task design, all of the stages of the workflow were designed to appear in front of the users for only a limited amount of time, usually a few minutes on average. Users could follow the timer at every step of the collaboration as indicated by a visible countdown event. Moreover, various time lengths were tested through a number of trials at development mode. The results helped establishing a reasonable timeline that would best accommodate time-on-task and user performance. The choice of restricting the work spaces to finite window frames comes from the need to meet the constraints of task synchronicity, time management and attention retention. This technique is closely related to the concepts of timeboxing \cite{jalote2004timeboxing} and iterative development used in software design and pair programming \cite{larman2003iterative}. The benefit of this sort of time management techniques \cite{cirillo2009pomodoro}, designed around enhancing performance and mono-tasking, is the known reduced impact of internal and external interruptions on focus and flow. It provides a clear structure and sustained focal point around the single priority at hand. Limiting time on task also tries to address the risk that extended periods of effort on a single task lead to performance decline \cite{kurzban2013opportunity}. As with every design choice, drawbacks also exist, and these are mainly related to the inherited limitations to thinking and planning time, as well as the need for users to effectively manage the collaboration space within fixed timelines. The collaborative task requires cooperation between teammates and quick judgment of their characteristics for team formation. Fitting together time management, complex decision making and time-on-task attention depletion into a functional collaborative framework is one of the greatest challenges met by our framework. The complexity of this challenge becomes greater as the number of workers increases. Therefore, to attain a fair level of scalability for our framework, we rely on concepts such as thin-slicing \cite{sridharan2007thin} and cognitive heuristics \cite{metzger2013credibility} to reduce the mental effort of the workers and aid the path to problem solving by confining the task allotment to fixed sizes which can be adjusted to the level of complexity of the given subject. The sample story, used in all of the experiments, was designed according to its simplicity, brevity and ambiguity of meaning, and purposefully crafted to trigger a number of diverse responses from the collaborators. Testing other sample stories and topics, as well as different lengths of the exercises could further the contribution of our research on team formation for creative and complex tasks, independently of the type of subject.

 \subsection{Competitive self-organisation and peer review as built-in quality assurance mechanisms}
A typical risk in online work settings is the low level of veracity in worker responses, e.g. in terms of the evaluation scores workers give to their teammates. As shown in \cite{gaikwad2016boomerang}, it is often difficult to distinguish good from low-performance workers in paid crowdsourcing platforms, because most workers have high reputation scores as a result of social pressure. Competitive self-organisation on which the SOT approach is based, is in line with the main idea championed in the aforementioned paper, i.e. the need to incentivize accurate feedback by rebounding the consequences of feedback to the person who gave it. By giving truthful responses in regards to their teammate's capabilities, and by selecting the people with who they really want to work with, workers give themselves more chances to collaborate with a good teammate, and thus to win. In contrast, if workers give high scores to all their teammates, regardless of actual skill or collaboration capability, their chances to be paired with an appropriate teammate -- and thus win -- are lowered. 
This competitive element, which is built directly into the SOT framework, becomes in this way a powerful drive of truthful behavior. As a second built-in quality mechanism, the proposed SOT framework uses peer assessment at the end of each round, in order to select the story fragment that will be used to continue the main story. As shown in~\cite{whiting2017crowd}, peer assessment is another powerful mechanism to ensure that the best outcome is chosen and allow the best capabilities of the worker collective to fully emerge.

\textit{The above two quality mechanisms, i.e. competitive self-organisation and peer assessment, complement the importance of the element of agency within the SOT framework}. Relying on the element agency alone could mean that workers choose teammates with whom they get along well, but not necessarily those with whom their performance is optimized, especially in the presence for example of social pressure. By adding the element of competition (primarily) and peer assessment (secondarily) workers have a strong motivation to accurately assess the capability of the person they choose.
Aside from the above two quality mechanisms, in the future, it would be interesting to investigate which additional mechanisms, for example in the form of intrinsic motivators, can be added to the SOT framework to further safeguard truthful responses and high quality of work.

\subsection{Extending to larger teams} \label{from_dyads_to_teams}
A long-term scholarly debate exists in small groups' literature on whether dyads constitute teams, which inspires our following discussion on applying the proposed model to triads or even larger groups. On the one hand, scholars such as~\citet{moreland2010dyads} are of the opinion that the size of a team should be at least three, because dyads are more ephemeral than larger groups, and certain phenomena like majority/minority relations, coalition formation, and group socialization can only be observed in larger groups. Other researchers, such as Williams~\cite{williams2010dyads}, disagree, arguing that two people can be considered to be a team, since 
some of the most interesting group processes, like inclusion/exclusion, power dynamics, leadership and followership, cohesiveness, social facilitation, and performance occur in dyads in the same manner that they do in larger groups, and that, in most instances, dyads operate under the same principles that explain group dynamics in teams of three or more. 

In the experimental part of this study we worked with dyads, as we are primarily interested in the team processes that  are present already from the dyad setup (such as inclusion/exclusion, leadership/followership, or performance), and less interested in phenomena that occur exclusively in large groups (like majority/minority relations). This choice was also motivated by the fact that in the particular domain of online collaboration, it is not uncommon to work with pairs as the essential foundations for studying team phenomena that have implications for larger groups~\cite{lykourentzou2017team,miller2014pair,chikersal2017deep,haake2004end,mcdonald2000expertise,nardi2006strangers,yarosh2013need}.

Nevertheless our proposed model can be extended quite straightforwardly to cover groups of three members or more. Specifically, the algorithm that helps form the teams first creates a complete graph with candidate team members as nodes and average pairwise ratings as edges. The algorithm then produces all possible graph cuts of a given team size, which in our experiments was set to two. Next, the algorithm greedily selects those teams that maximize inter-group affinity, eliminating all alternative teams that the selected individuals could have participated into, until all individuals belong to a team. The size of the cut is a parameter, and the algorithm can be adapted to compute all possible cuts of a given size, equal to the size of the desired teams (naturally, depending on the specified team size the teams formed last may have less members). The specific algorithm is greedy, as we note that the problem of optimally partitioning a complete graph into subsets of equal cardinality falls under the NP-Hard complexity category~\cite{feldmann2015balanced}. Other approaches, including heuristics and metaheuristics, can also be used to create teams efficiently, for example an adaptation of the k-nearest neighbor algorithm. Finally, in case the goal is to specify the number of teams to be created, rather than the team size, polynomial-time algorithms can adapted, such as the algorithm proposed by Andreev and Racke to solve the Balanced Graph Partitioning problem~\cite{andreev2006balanced}.

\subsection{Theoretical and design implications for future system design}

Our proposed model and system has a number of implications for future system design, particularly for the design of online work platforms. We discuss these in the following.

A first implication concerns the advantages and disadvantages of purely member-driven teams, and, consequently, the desired level of algorithmic involvement. Our work is driven by the need to mitigate the problems brought by the tight algorithmic supervision of current top-down team formation systems, which range from inefficient collaboration to significant psychological discomfort. As we also show in this paper, incorporating agency can improve the outcomes of collaborative work and worker well-being. In the long run, giving workers more control of whom they work with and how can help create online work systems that are more empowering and offer more opportunities for personal development for the participating workers. Despite its many advantages, user agency in teammate selection also comes with ethical concerns and potential risks for the workers, but also for the platforms. Delegating team formation fully to human participants means that some workers may be more sought after than others. This can prove beneficial for the collaboration (e.g., a person knowing that they can work better with a certain teammate), but it may also be detrimental (e.g., exclusion of specific individuals due to demographic factors, similarity, or familiarity~\cite{hinds2000choosing}). Recent evidence from the Human-Computer Interaction community~\cite{gomez2019would}, shows that allowing full individual control may be at least as problematic because it can replicate systematic inequality, exclude people who do not ``look like’’ good team members, or lead to segregated teams. To mitigate possible selection biases, our system can be expanded to explicitly promote diversity. For example, in an approach inspired by traditional recommender systems~\cite{ kunaver2017diversity}, the SOT algorithm could be parameterized to promote candidate teammate profiles that the user has not seen or selected before, based on feature dissimilarity. Alternatively, the system could be parameterized to reward (monetarily, or with more time for example) workers for collaborating with people outside their ``comfort zone’’. Another potential risk stemming from fully human-led team formation is the exclusion of low-ranked users, for example those that received low scores in early evaluations, or newcomers (for versions of the system that allow this). To help those users recover and avoid segregation, the SOT algorithm could be parameterized to include the element of serendipity, or to explicitly reward ``mentorship’’, i.e. teams that mix low-ranked and high-ranked users. Finally, recommendation diversity and hierarchical clustering techniques can prove useful to ensure the scalability of the system. In our experiments we used batches of up to 12 people, however batches with more workers may be available or necessary for a particular task. To help workers efficiently process dozens or hundreds of candidate worker profiles, one could envision extending our proposed SOT algorithm with hierarchical clustering recommendations, which start by presenting to the worker diverse teammate ``types'', and then allow the user to explore clusters of candidate teammates that are similar to their preferred type. What is important to note here is that, it is in such situations, where algorithmic involvement can prove beneficial to the teammate selection process rather than blocking it.

Another risk of full self-organization, this time impacting the whole online work system, is platform disintermediation in which workers negotiate, collaborate, and transact with one another, and potentially with the client, outside the platform boundaries. \textit{Constrained self-organization}, such as the one advocated in our proposed system, where an algorithm assists worker teammate negotiations, has been conceptually proposed~\cite{ jarrahi2020platformic} as the golden mean between safeguarding worker autonomy and protecting digital work platforms from disintermediation.

Another design implication comes from the form used by the system to grant agency to the workers. One can distinguish two main methods of granting user agency: direct negotiations and mediated approaches. Direct agency negotiations take place when the system enables users to ask others whether they want to become teammates, be able to answer positively or negatively, and directly negotiate with several candidate teammates until the groups are formed~\cite{guimera2005team, zhu2013motivations}. The advantage of this approach is that it allows workers to explore whether a collaboration should proceed and why fully. The disadvantage is that the negotiation process can be lengthy, and the number of possible collaborations that can be explored is minimal due to the human cognitive limitations of processing rich information about one’s candidate teammates (including verbal and also non-verbal communication, e.g., through video), as well as the fatigue that inevitably accompanies this process. 
Another disadvantage of direct negotiation comes from exposing users to the personal disclosure of explicitly declaring their interests and rejections towards their teammates. In popular person-to-person recommender systems \cite{david2016screened} the user first states their intent at the system level (e.g. dating apps permitting users to swipe left/right) before allowing the newly matched pair to establish unmediated communication.
Mediated agency approaches, such as the ones explored in~\cite{tacadao2015generic, manya2017maxsat, meulbroek2019forming} elicit users’ teammate preferences, through methods such as the maximum satisfiability problem (MaxSAT) \cite{hansen1990algorithms}, the comprehensive assessment for team-member effectiveness (CATME) \cite{barrick1998relating}, and the Gale-Shapley Algorithm \cite{gale1962college}. The advantages of this approach are that it allows workers to explore significantly more candidate teammates since the profile information to be processed per teammate is more concise and faster. The disadvantage is that it grants less time for reflection on a candidate teammate's suitability, and to a certain extent, it is also dependent on algorithmic involvement. 
The approach adopted in this work fits more in the second category, i.e., mediated agency, since the SOTs system uses the participants’ explicit stated (and not deduced) preferences about their teammates, rather than the outcome of their in-between direct negotiations. This design choice is more appropriate for online and crowd work platforms, where the tasks are time-bounded, and there is a need to maintain scalability in the presence of dozens of candidate teammate profiles to choose from. However, at the same time, our system allows users to directly ``negotiate'' with one another, in the sense that workers can try out different teammates across the rounds by actually working with them. 
Future design frameworks could explore how direct negotiations could be better incorporated into large-scale online work systems without negatively impacting the task's completion. One possible solution in this direction could be combining direct and mediated negotiations in individual tasks, and then maintaining a database of the negotiation results across the tasks, which can be progressively enriched by the workers as they discuss with more teammates over time.

\subsection{Limitations}
This work also has a number of limitations. Task type is one of them. In this study we examined a particular task, 
i.e. fictional story writing. This task belongs to the broader category of complex, open-ended and creative tasks, which require giving workers full task context (in this case about the main story), and which cannot be easily decomposed to micro-task level. Tasks of this type are not routinely handled through crowd work but are nevertheless crucial to our knowledge economy, for solving problems that range from creative idea generation to creative disaster response, to name just a few. 
Although the particular task type does not require specialized skills, it does allow distinguishing between different levels of creativity and prior experience specific to the task (in this case in story writing) that the workers may possess in various degrees. In this sense, the task allows participants to filter their candidate co-workers based on their creative skills, and in parallel it is suitable for the workforce available in commercial crowd work platforms.
 
In the future it would be useful to explore self-organization for other types of tasks that are more frequently encountered in online work settings, such as those requiring stricter workflows, those with predefined team roles, tasks requiring expert skills or those that can be decomposed to smaller work units. Self-organization in those contexts could mean exploring how workers are given agency to not only choose their teammates, but how to split work responsibilities and delegate task parts among their teammate circles, in line with latest directions on crowd work reported in the literature~\cite{wood2019networked}.

Another limitation 
is the degree of agency given to the workers, and the fact that worker agency is restricted to teammate choice. Although this is one of the first studies to explore self-organization in an online crowd work context, it did limit worker agency to the decision about which teammates to work with. The rest of the workflow settings, like the timing of the activities, the number of rounds, size of the worker batch, etc., were not up to the workers to decide, and this inevitably poses limits on their self-management potential. Future work could explore the relative weight of the aforementioned workflow design settings, in the final result, through a series of experiments, each focused on a particular setting. Future work could also explore giving workers \textit{agency over the entire workflow design}, in order to better tailor their collective work strategy to the needs of the particular task.

Finally, the design used in this study combines the elements of competition and collaboration, which are often seen in real-life crowdsourcing scenarios (e.g., Kaggle competitions). We retained the element of competition and its surrogate product of user popularity as it is a secondary result of rating users for their output. Users in our study can also see their peers' expertise information, similarly to how people almost always become aware of the expertise of other teammates in real-world team applications. Nevertheless popularity can also impact teammate choice beyond the necessities of the task, and into selection biases. To mitigate this, future systems could offer the option to reveal the users’ performance, or any other popularity scores, as an opt-in/opt-out component depending on the scope of the system.

\section{Conclusion}
This paper investigates the effects of a novel online team formation framework, titled Self-Organizing Teams(SOTs). SOTs places increased emphasis on user agency over team formation and relies on the collective decisions of online workers to self-organize into effective teams, while being supported -- but not guided -- by an algorithm. 
We compared the SOT framework with two baselines, where individuals are allocated the same teammate throughout a creative online task, either with the illusion of user agency (placebo condition) or without it (no-agency condition). Allocating the same teammate for the entire task duration is a typical approach of forming teams in online work situations. Our findings indicate that the SOTs method leads to a higher quality output, as measured by independent evaluators. 
Furthermore, we carried out a set of quantitative analysis of the worker's perception of the collaboration, which showed that teams formed under the SOT condition are more satisfied during their collaboration, and able to collaborate and help each other more.

The purpose of this paper is to lay the ground for the analysis of crowd-led collaboration aided by non-intrusive models, which are inspired by bottom-up self-organizing systems. It is our hope that this work will inspire more researchers to look into online team formation systems that move away from the trend of micro-managing workers, and into the direction of making the latter an integral part of the algorithmic decision-making process. In the future, we aim to integrate more complex user-preference elicitation methods into the approach, for example by using explanations to allow users to collectively set the objective function weights and decide which team formation elements are most important to solve the complex problem at hand. As an example of the above, workers could use the SOTs framework to decide how much diversity or homogenerity a certain task requires, and balance this with the need to optimise for skill or personality complementarity.

\bibliographystyle{ACM-Reference-Format}
\bibliography{main}

\appendix

\section{Annex}

\begin{table}[h!]
\centering
\caption{The story quality ratings by the external evaluators were significantly correlated at the 0.01 level (2-tailed), N= 1960.}
\label{table:pearson_external_eval}
\begin{tabular}{ |c|c|c|c|c|c| } 
 \hline
 \multicolumn{6}{|c|}{Pearson Correlations} \\
 \hline
        & Grammar & Interest & Originality & Plot & Overall \\ 
Grammar & 1 & $.537^{**}$ & $.448^{**}$ & $.559^{**}$ & $.589^{**}$\\ 
Interest & $.537^{**}$ & $1$ & $.617^{**}$ & $.753^{**}$ & $.809^{**}$ \\
Originality & $.448^{**}$ & $.617^{**}$ & 1 & $.550^{**}$ & $.601^{**}$ \\
Plot & $.559^{**}$ & $.753^{**}$ & $.550^{**}$ & 1 & $.739^{**}$ \\
Overall & $.589^{**}$ & $.809^{**}$ & $.601^{**}$ & $.739^{**}$ & 1 \\
\hline
\end{tabular}
\end{table}

\end{document}